\documentclass[journal,10pt,compsoc]{IEEEtran} 
\usepackage{fix-cm}
\IEEEoverridecommandlockouts
\usepackage{bm,bbm,amsthm,amsmath,amssymb,amsfonts,mathrsfs}
\usepackage[ruled,vlined]{algorithm2e}
\usepackage{textcomp}
\usepackage{xcolor}
\usepackage{graphicx}  
\usepackage{float}  
\usepackage{subfigure}  
\usepackage{colortbl,booktabs,threeparttable}  
\usepackage{array}
\usepackage{booktabs}
\usepackage{boldline}
\usepackage{url}  
\usepackage{color}
\usepackage{eqparbox}
\usepackage{caption}
\usepackage{stfloats}
\usepackage{verbatim}
\usepackage{multirow,booktabs,threeparttable}

\usepackage{siunitx}
\usepackage{balance}

\def\BibTeX{{\rm B\kern-.05em{\sc i\kern-.025em b}\kern-.08em
    T\kern-.1667em\lower.7ex\hbox{E}\kern-.125emX}}

\newtheorem{definition}{Definition} 
\newtheorem{assumption}{Assumption} 
\newtheorem{theorem}{Theorem}

\newtheorem{lemma}{Lemma}

\usepackage{subcaption}
\usepackage[bookmarks=false,colorlinks,citecolor=black, filecolor=black, linkcolor=black,urlcolor=black]{hyperref}
\usepackage{cleveref}
\allowdisplaybreaks[4]

\DeclareMathOperator*{\argmin}{arg\, min}

\begin{document}
\title{Multi-agent Uncertainty-Aware Pessimistic Model-Based Reinforcement Learning for Connected Autonomous Vehicles}

\author{Ruoqi Wen, Rongpeng Li, Xing Xu and Zhifeng Zhao
    \thanks{R. Wen and R. Li are with the College of Information Science and Electronic Engineering, Zhejiang University, Hangzhou 310058, China (email: \{wenruoqi, lirongpeng\}@zju.edu.cn).}
    \thanks{X. Xu is with the Information and Communication Branch of State Grid Hebei Electric Power Co., Ltd, China (e-mail:hsuxing@zju.edu.cn).}
    \thanks{Z. Zhao is with Zhejiang Lab, Hangzhou 311121, China, and also with the College of Information Science and Electronic Engineering, Zhejiang University, Hangzhou 310058, China (email: zhaozf@zhejianglab.com).}
}
\maketitle

\begin{abstract}
Deep Reinforcement Learning (DRL) holds significant promise for achieving human-like Autonomous Vehicle (AV) capabilities, but suffers from low sample efficiency and challenges in reward design. Model-Based Reinforcement Learning (MBRL) offers improved sample efficiency and generalizability compared to Model-Free Reinforcement Learning (MFRL) in various multi-agent decision-making scenarios. Nevertheless, MBRL faces critical difficulties in estimating uncertainty during the model learning phase, 
thereby limiting its scalability and applicability in real-world scenarios. Additionally, most Connected Autonomous Vehicle (CAV) studies focus on single-agent decision-making, while existing multi-agent MBRL solutions lack computationally tractable algorithms with Probably Approximately Correct (PAC) guarantees, an essential factor for ensuring policy reliability with limited training data. To address these challenges, we propose \texttt{MA-PMBRL}, a novel Multi-Agent Pessimistic Model-Based Reinforcement Learning framework for CAVs, incorporating a max-min optimization approach to enhance robustness and decision-making. To mitigate the inherent subjectivity of uncertainty estimation in MBRL and avoid incurring catastrophic failures in AV, \texttt{MA-PMBRL} employs a pessimistic optimization framework combined with Projected Gradient Descent (PGD) for both model and policy learning. \texttt{MA-PMBRL} also employs general function approximations under partial dataset coverage to enhance learning efficiency and system-level performance. By bounding the suboptimality of the resulting policy under mild theoretical assumptions, we successfully establish PAC guarantees for \texttt{MA-PMBRL}, demonstrating that the proposed framework represents a significant step toward scalable, efficient, and reliable multi-agent decision-making for CAVs.
\end{abstract}


\section{Introduction}
\IEEEPARstart{R}{esearch} on Connected Autonomous Vehicles (CAVs) has gained considerable momentum recently, particularly concerning developing advanced control algorithms \cite{zhang2024survey, kiran2021deep}. Multi-Agent Reinforcement Learning (MARL) has emerged as a promising approach for enabling CAVs to execute complex tasks autonomously. Traditionally, Model-Free Reinforcement Learning (MFRL), which leverages observed rewards from state-action transitions, has been widely adopted in this field \cite{MAPPO, FIRL, SVMIX}. However, the costly requirement for sufficient data through extensive real-world interactions makes MFRL stuck in unstable learning and high computational overhead, thus making it less competent in autonomous driving scenarios. On the other hand, most multi-agent algorithms in the field of vehicular networks have focused on the Centralized Training and Decentralized Execution (CTDE) framework \cite{10.5555/3305381.3305500}. Despite its strong performance in some scenarios, the centralized training approach encounters limited scalability and suffers severe performance degradation in communication-constrained or unreliable environments \cite{ma2024efficient}. To tackle these challenges, some studies have explored fully Decentralized Training and Decentralized Execution (DTDE) algorithms \cite{matignon2012independent, lin2021multi} 
and adopted Model-Based Reinforcement Learning (MBRL)  \cite{wu2022uncertainty,pan2021integrated,ma2024efficient} for scenarios with communication constraints.

In contrast to MFRL, MBRL \cite{wu2022uncertainty,pan2021integrated,ma2024efficient} typically reduces the need for real-world interactions by building an environmental dynamics and reward model, which facilitates subsequent training, minimizing reliance on the real world, and improving adaptability and generalization to dynamic changes. 
Meanwhile, arising from limited data and inherent system noise, inaccuracies in the learned model can significantly compromise the stability and robustness of subsequently learned policies, especially in methods like Model Predictive Control (MPC) \cite{camacho2013model}, where errors in state transitions and future predictions accumulate \cite{nagabandi2018neural, PETS}. Therefore, there emerge some modified Dyna-style algorithms \cite{10.1145/122344.122377} to address model inaccuracies and enhance policy performance. For example,  \cite{wu2022uncertainty} introduces an uncertainty-aware MBRL algorithm \texttt{UA-MBRL}, which utilizes model ensemble techniques to quantify output uncertainty and introduces an adaptive prediction truncation method to reduce potential performance degradation. In addition to the uncertainty estimation problem, MBRL faces a significant Out-Of-Distribution (OOD) issue when the dataset is limited, hindering most algorithms from learning better policies without sufficient data support. To address this, existing algorithms similar to \texttt{UA-MBRL} \cite{wu2022uncertainty}, including \texttt{MOPO} \cite{NEURIPS2020_mopo} and \texttt{MORel} \cite{NEURIPS2020_morel}, typically rely on idealized assumptions regarding uncertainty estimation. For instance, uncertainty quantification techniques such as truncation \cite{wu2022uncertainty} inherently introduce subjectivity, leading to cumulative errors and model bias. Furthermore, explicit uncertainty estimates used by these approaches can be unreliable, particularly when derived from Deep Neural Network (DNN)-based models \cite{uehara2022pessimistic}. Therefore, shifting toward more robust algorithms that do not depend on explicit uncertainty estimation can help overcome the inherent limitations of uncertainty quantification.

Robust MBRL \cite{wang2021online, SWAZINNA2021104366} aims to optimize worst-case performance over an uncertainty set of Markov Decision Processes (MDPs). 
Incorporating perturbations during training can improve agent stability, generalization, and performance, especially in safety-critical domains such as CAV scenarios. For instance, \cite{uehara2022pessimistic} introduces the \texttt{CPPO} algorithm, a model-based pessimistic offline RL method that incorporates \emph{pessimism} by searching for the worst-case transition model within a predefined constraint set and optimizes the policy through a max-min operation. Utilizing general function approximation, \texttt{CPPO} establishes an upper bound on the sub-optimality gap under partial coverage assumptions and offers Probably Approximately Correct (PAC) guarantees. Notably, partial coverage bypasses the challenging assumption that the density ratio-based concentrability coefficient is bounded for \emph{all} learned policies. Instead, it only assumes the boundness of the optimal policy \cite{uehara2022pessimistic}. However, despite its theoretical contributions, the algorithm provides only a conceptual framework that is challenging to implement in practice. Therefore, based on the theoretical contributions of \texttt{CPPO} \cite{uehara2022pessimistic}, \cite{ramborl} develops a computationally efficient algorithm \texttt{RANMBO-RL}, which enhances robustness by generating pessimistic synthetic transitions for OOD data and reformulating the policy optimization problem as a two-player zero-sum game against an adversarial environment model. Additionally, \cite{hongpessimistic} proposes a potential algorithm to implement a pessimistic framework, but it lacks empirical validation and is limited to the single-agent scenario. However, the extension to a multi-agent case with theoretical PAC guarantees, which are essential for ensuring policy reliability when training data is limited, remains unexplored, and the challenge is further complicated by the absence of computationally tractable multi-agent MBRL algorithms.

%

In this paper, we investigate self-decision control in CAVs for intelligent traffic management within a fully DTDE framework, tailored explicitly for scenarios with communication constraints and limited datasets with partial coverage. To deliver a cost-effective MARL solution with theoretical guarantees, we propose a novel decentralized algorithm, termed Multi-Agent Pessimistic MBRL (\texttt{MA-PMBRL}), on top of Soft Actor-Critic (SAC) \cite{SAC}. Specifically, \texttt{MA-PMBRL} facilitates the efficient exchange of collected samples among individual agents and their neighboring vehicles within a specified communication range. In addition, it incorporates a pessimistic model-based min-max framework, which ensures vehicle safety by mitigating the reliance on overly subjective uncertainty estimations. This enhancement ensures vehicle safety, albeit with a potential trade-off in performance.
Furthermore, unlike \texttt{RANMBO-RL}, which addresses the pessimistic min-max problem by reformulating it as a two-player zero-sum game against an adversarial environment model, \texttt{MA-PMBRL} leverages Projected Gradient Descent (PGD) \cite{jain2017non} within the critic network. This approach greatly simplifies the computational process of deriving effective control policies from pessimistic estimations.
While highlighting the key difference with the existing literature in Table \ref{tb: parameters}, the contributions of this work can be summarized as follows:
\begin{itemize}
\item 
We propose a sample-efficient and computation-implementable Multi-Agent Pessimistic MBRL algorithm \texttt{MA-PMBRL}, built upon SAC and a max-min framework to enhance performance and robustness. \texttt{MA-PMBRL} effectively avoids subjective uncertainty estimation in MBRL without compromising performance while leveraging a fixed dataset with partial coverage to address covariate shifts in MBRL.
\item 
Building upon the theoretical foundations of graph theory \cite{orlin1977contentment} and concentration inequalities \cite{Boucheron2004}, we provide a rigorous theoretical analysis of \texttt{MA-PMBRL} under communication constraints and partial coverage. Unlike \cite{uehara2022pessimistic} and \cite{hongpessimistic}, our work provides a more comprehensive comparative analysis for multi-agent settings. Specifically, we enhance and refine its theoretical derivations, addressing previously incomplete aspects of the original proofs. Additionally, our analyses reveal how the communication range affects the convergence of MARL, which is further rigorously validated in challenging urban driving scenarios. Notably, unlike common MFRL methods, our result does not rely on the assumption of Bellman completeness \cite{NEURIPS2021_34f98c7c}. 
\item 
Within the multi-agent framework, we also propose a communication protocol under communication constraints, thus facilitating the efficient exchange and better balancing communication overhead and system performance. Extensive simulations validate the superior sample efficiency of \texttt{MA-PMBRL} over other solutions like \texttt{MA-PETS}\cite{wen_multiagent_2024}, \texttt{SVMIX} \cite{SVMIX}, \texttt{FIRL} \cite{FIRL}, \texttt{MBPO} \cite{MBPO}, \texttt{SAC} \cite{SAC}.
\end{itemize}
\begin{table}
    \caption{\centering{The Key Parameters Used in the Paper.}}
    \begin{center}
    \label{tb: parameters}
            \begin{tabular}{|c|m{6.8cm}|}
                \toprule	
                \multicolumn{1}{|c|}{\textbf{Notation}}& \multicolumn{1}{c|}{\textbf{Parameters Description}} \\
                \midrule
                $I$& Number of RL agents \\\hline
                $K$ & No. of episodes  \\\hline
                $H$ & Length per episode  \\\hline
                $\gamma$ & Discount factor   \\\hline
                $\mathcal{M}_\mathcal{D}$ & Constraint model class driven by $\mathcal{D}$\\\hline
                 $\mathcal{D}_E$ & Buffer for storing actual data \\\hline
                 $\mathcal{D}_{\tilde{T}_\phi}$ & Buffer for storing virtual data \\\hline
                 $T$ & True dynamics model of the unknown environment\\\hline
                 $T_\phi$ & Candidate model of constraint model class $\mathcal{M}_\mathcal{D}$\\\hline
                 $\tilde{T}_\phi$ & Solution of the pessimistic model-based min-max RL problem\\\hline
                 $L_{\text{rollout}}$ & Rollout length   \\\hline
                 $X$ & Maximum iteration of PGD updates \\\hline
                 $\phi_\mathrm{best}$ & Smallest objective value of PGD\\\hline
                $d$ & Communication range \\\hline
                $\mathcal{B}_{t}^{(i)}$ & The set of neighboring CAVs for vehicle $i$ at $t$\\\hline 
                $v_t$ & Speed per vehicle at time-step $t$ \\\hline
                $z_t$ & Position per vehicle at $t$  \\\hline
                $\bar{v}_{t}$ & Target velocity per vehicle at $t$  \\\hline
                $v_{t,a},v_{t,e}$ & Speed of vehicles \underline{a}head and b\underline{e}hind  \\\hline
                $l_{t,a},l_{t,e}$ & Distance of vehicles \underline{a}head and b\underline{e}hind \\\hline
                $b^{(i)}$ & Vehicle in emergency braking range (1: yes, 0: no)\\\hline
                ${\Lambda}x_t$ & Travel distance per vehicle at $t$  \\\hline
                $C_{\pi^\ast}$ & Concentrability coefficient for $\pi^\ast$ \\\hline
                $\Delta_a$ & Performance gap between the true model and estimated model under $\pi^\ast$ \\\hline
                $\Delta_b$ &Discrepancy between the best constraint model class under $\pi^\ast$ and $\pi_t$ \\\hline
                $\Delta_c$ &Performance gap between the best constraint model class and the true model under $\pi_t$ \\\hline
                $\Gamma_a$ &Expected advantage function difference under the best constraint model class\\\hline
                $\Gamma_b$ &Expectation of the weighted gradient of the policy log-likelihood under the best constraint model class\\\hline
                $\Gamma_c$ &Difference in expected policy gradient terms under the best constraint model class\\\hline
                $c_1$ & Scaling constant associated with approximation and concentration bounds\\\hline
                $c_2$ & Adjustment factor for complexity term, influencing worst-case error estimation at confidence level $1-\delta$ \\\hline
                $c_3$ & Upper bound on the maximum weighted feature $w$ and influence of feature $\varphi$ \\\hline
                $\mathbf{C}_{d}$ & Cliques cover of $\mathcal{G}_{d}$\\\hline
                $\bar{\chi}\left(\mathcal{G}_{d}\right)$ & Minimum number of cliques cover  \\
                \bottomrule
            \end{tabular}
    \end{center}
\end{table}
The remainder of the paper is organized as follows. Section \ref{sec: preliminaries} briefly introduces the mathematical preliminaries and formulates the problem to be investigated. In Section \ref{sec: MA-PMBRL} and in Section \ref{sec: convergence proof}, we provide more details of \texttt{MA-PMBRL} and establish an upper bound for the sub-optimality of its yielded policy, respectively. Finally, Section \ref{sec: numerical experiments} demonstrates the effectiveness and superiority of \texttt{MA-PMBRL} through extensive simulations. We conclude the paper in Section \ref{sec: conclusion}.

\section{Related Works}
\label{sec: related works}
\subsection{Multi-Agent Reinforcement Learning for CAVs}
The complexities of urban high-density mixed-traffic intersections present both challenges and opportunities to improve traffic efficiency and safety through CAV interactions, with MARL garnering attention for its potential to enhance these aspects \cite{9762548, 9939107, Hua2023MultiAgentRL}. Incorporating communication, such as Vehicle-to-Vehicle (V2V) interaction, improves coordination, reduces collisions, and enhances decision-making \cite{banjanovic2016autonomous}. However, challenges in scalability, stability, and managing non-stationary environments persist. In this regard, CTDE-based approaches enable coordination during training through information sharing while maintaining independent action during deployment. For instance, within the actor-critic MFRL framework, \cite{MADDPG} introduces \texttt{MA-DDPG}, which extends \texttt{DDPG} to multi-agent settings with CTDE. Additionally, \cite{foerster2018counterfactual} proposes \texttt{COMA}, which improves credit assignment through counterfactual experience replay, enhancing multi-agent collaboration. However, as the number of agents or the action space grows, the computational complexity of MFRL algorithms can increase exponentially, posing a significant challenge. 
To reduce the computational burden, \cite{SVMIX} proposes using a stochastic graph neural network to capture dynamic topological features of time-varying graphs, while decomposing the value function. To address the communication overhead caused by frequent information exchanges, \cite{FIRL} introduces a consensus-based optimization approach built on periodic averaging, integrating the consensus algorithm into Federated Learning (FL) for local gradient exchange.

While these studies highlight the strong post-training performance of RL in real-time applications, the MFRL algorithm remains hindered by high overhead due to its significant computational and sampling demands. This challenge is especially evident in scenarios like CAVs \cite{drl_AV}, where data acquisition is difficult and extensive agent-environment interactions are costly. As a result, with limited data samples, MFRL performance can rapidly deteriorate and become unstable.

\subsection{Model-Based Reinforcement Learning}
MBRL naturally emerges as an alternative solution to address the sampling efficiency and communication overhead issues in MFRL. Unfortunately, MBRL suffers from performance deficiency, as it might fail to estimate the environmental uncertainty accurately and characterize the dynamics model, a critical research component in MBRL. For example, PILCO~\cite{deisenroth2011pilco} marginalizes the aleatoric and epistemic uncertainty of a learned dynamics model to optimize the expected performance. Most MBRL algorithms suffer from subjectivity in uncertainty estimation. Taking the example of the variants of \texttt{MBPO} \cite{MBPO}, \cite{NEURIPS2020_mopo} introduces model-based offline policy optimization (\texttt{MOPO}), which optimizes the policy under the uncertainty-penalized MDP framework. By incorporating a soft reward penalty based on model error estimates, the policy can return to high-confidence regions after taking risky actions, without being terminated. Meanwhile, \cite{NEURIPS2020_morel} introduces \texttt{MoRel}, which defines terminal states by combining a subjective threshold based on uncertainty and uses a large negative reward to penalize unknown regions, thereby enhancing the robustness of decision-making in RL.

To address subjective uncertainty estimation and OOD issues, \cite{uehara2022pessimistic} proposes \texttt{CPPO}, which introduces a constrained pessimistic policy optimization to limit excessive optimization and improve stability in uncertain environments from offline data with partial coverage. Building on this, \cite{ramborl} introduces \texttt{RAMBO-RL}, incorporating a two-player zero-sum game formulation against an adversarial environment model, where the model is trained to minimize the value function while ensuring accurate transition predictions. Additionally, \cite{MAPPO} introduces the \texttt{MAPPO} algorithm, which integrates a prior model into the \texttt{PPO} framework to accelerate the learning process using current centralized coordination methods. Despite relying on local models, these methods are either single-agent algorithms or depend on a global critic. However, centralized coordination methods suffer from high resource consumption, lack of flexibility, and delays, as the central node processes vast amounts of information for the entire system, limiting effectiveness in communication-constrained environments \cite{wu2023models}.
Additionally, CTDE relies on global information during training, which is challenging to implement in real-world applications, and faces scalability issues in large-scale or resource-constrained deployments. Although some recent works attempt to adopt fully decentralized MBRL algorithms (e.g., multi-agent probabilistic ensembling and trajectory sampling \cite{wen_multiagent_2024}, decentralized policy optimization and agent-level topological decoupling \cite{ma2024efficient}) in communication-limited vehicular networks, limited endeavors exist to solve the problem of overly subjective uncertainty estimation in multi-agent MBRL \cite{survey_mbrl}. On the other hand, considering the substantial communication overhead inherent in centralized coordination methods, we focus primarily on developing a fully decentralized multi-agent MBRL algorithm that operates efficiently under communication constraints.

\section{Preliminaries and System Model}
\label{sec: preliminaries}
This section briefly introduces some fundamentals and necessary assumptions of the underlying MDPs and one framework of single-agent pessimistic MBRL. Regarding the decision-making issue for CAVs, we highlight how to formulate the problem under the multi-agent case. 

\subsection{Preliminaries}
\subsubsection{Parallel Markov decision process}
The decision-making problem of CAVs can be formulated as a collection of \textit{parallel time-homogeneous stochastic MDPs} $\mathcal{M}:= \left\{M\left(\mathcal{S}^{(i)}, \mathcal{A}^{(i)}, \mathcal{P}^{(i)}, \mathcal{R}^{(i)}, \mu_0, \gamma, H\right)\right\}_{i=1}^I $ among agents $i \in [I]:=\{1,2,\cdots, I\}$ for $H$-length episodes \cite{bernstein2002complexity}. Despite its restrictiveness, parallel MDPs provide a valuable baseline for generalizing to more complex environments, such as heterogeneous MDPs. Notably, agents have access to identical state and action space (i.e., $\mathcal{S}^{(i)}=\mathcal{S}^{(j)}$ and $\mathcal{A}^{(i)}=\mathcal{A}^{(j)}$, $\forall i, j \in[I]$), $\mathcal{R}^{(i)}: \mathcal{S}^{(i)}\times\mathcal{A}^{(i)}  \rightarrow [- R_{\max}, R_{\max}]$ is the reward function, $\mathcal{P}^{(i)}: \mathcal{S}^{(i)}\times\mathcal{A}^{(i)}\times\mathcal{S}^{(i)} \rightarrow \mathbb{R}^{+}$ is the transition functions, $\mu_0$ is the initial state distribution, and $\gamma$ the discount factor. Furthermore, we focus exclusively on stationary policies, denoted as $\pi^{(i)}: \mathcal{S}^{(i)} \rightarrow \mathcal{A}^{(i)}$, which 
represents a decision strategy to pick an action $a_t^{(i)}\in\mathcal{A}^{(i)}$ with probability $\pi^{(i)}(\cdot|s, a)$ given the current state $s_t^{(i)} \in \mathcal{S}^{(i)}$. 

In an MDP $M^{(i)}$, the standard objective is to identify a policy $\pi^{(i)}$ that maximizes the expected cumulative discounted reward, denoted as $V_{\mathcal{P},\pi}^{(i)}$ which is computed by integrating over the state space $\mathcal{S}^{(i)}$ under the transition dynamics $\mathcal{P}^{(i)}$, with an initial state distribution $\mu_0$ and a reward function $\mathcal{R}^{(i)}$ , and is formally defined as:
\begin{align}
V_{\mathcal{P},\pi}^{(i)} := \mathbb{E}_{\mathcal{P}, \pi ,s_0 \sim \mu_0}\left[\sum_{t=0}^{\infty} \gamma^t \mathcal{R}^{(i)}\left(s^{(i)}_t, a^{(i)}_t\right) \mid s^{(i)}_0\right],\nonumber
\end{align} 
where $\gamma$ is the discount factor, and $s_t^{(i)}$ and $a_t^{(i)}$ are the state and action at time step $t$, respectively.

In parallel MDPs, the state-action value function, $Q_{\mathcal{P},\pi}^{(i)}$, captures the expected cumulative discounted reward when starting from state $s^{(i)}_0$ and taking action $a^{(i)}_0$ at the first time step, followed by a policy $\pi^{(i)}$ thereafter, defined as:
\begin{align}
    Q_{\mathcal{P},\pi}^{(i)}:=\mathbb{E}_{\mathcal{P}, \pi,s_0 \sim \mu_0}\left[\sum_{t=0}^{\infty} \gamma^t \mathcal{R}^{(i)}\left(s^{(i)}_t, a^{(i)}_t\right) \mid s^{(i)}_0, a^{(i)}_0\right].\nonumber
\end{align} 
 
\subsubsection{Pessimistic model-based min-max reinforcement learning}
Common MBRL approaches typically utilize an estimated model of the MDP to assist in training a policy, since both the transition and reward functions are generally unknown. At the end of each episode, a fixed dataset $\mathcal{D}:=\left\{\left(s_t, a_t, r_t, s_t^{\prime}\right)_{t=1}^n\right\}$ can be used to infer the dynamics model and reward function through methods like Maximum Likelihood Estimation (MLE). 
For simplicity of representation, we write $T$ to unanimously denote both the learned dynamics and reward function, such as $T(s^\prime, r|s, a)$ is the probability of receiving reward $r$ and transitioning to $s^\prime$ after executing $(s, a)$. Therefore, for a true MDP $M=\left(\mathcal{S},\mathcal{A}, T, \mu_0, \gamma, H\right)$, the estimated one $\tilde{M}=\left(\mathcal{S},\mathcal{A}, \tilde{T}, \mu_0, \gamma, H\right)$ has the same state and action space, but uses the learned transition and reward functions. Notably, we assume that $\tilde{T}(\cdot|s, a)$ can be accurately parameterized by some $\phi$. 

Afterward, the goal of MBRL is to find the best possible policy contingent on the model learning from a fixed dataset, that is, $\tilde{\pi}=\arg \max _{\pi \in \Pi} V_{\tilde{M}}$.
To minimize reliance on assumptions about uncertainty estimation, we try to find the policy $\tilde{\pi}$ by solving a constrained max-min optimization formulation \cite{uehara2022pessimistic} as
\begin{align}
\label{eq:p_model_based_maxmin_rl}
    \tilde{\pi}=\underset{\pi \in \Pi}{\arg \max } \min _{T_\phi \in \mathcal{M}_{\mathcal{D}}} V_{T_\phi,\pi,\mu_0}, 
\end{align}
subject to the fixed dataset $\mathcal{D}$-driven constraint set of $\mathcal{M}_{\mathcal{D}}$
\begin{align}
\label{eq: constraint_MDP_original}
    \mathcal{M}_{\mathcal{D}}=\left\{T_\phi \Big\vert \mathbb{E}_{\mathcal{D}}\left[\mathrm{TV}\left(\tilde{T}_{\mathrm{MLE}}(\cdot \mid s, a), T_\phi(\cdot \mid s, a)\right)^2\right] \leq \xi\right\},
\end{align}
where $\mathrm{TV}(P_1, P_2)=\frac{1}{2} \sum_{(s, a,s^{\prime})\in\mathcal{D}}|P_1\big(s^{\prime} \mid s, a)-P_2(s^{\prime} \mid s, a)\big|$ represents the Total Variation (TV) distance \cite{devroye2018total} between distributions $P_1$ and $P_2$, while $\tilde{T}_{\mathrm{MLE}}$ denotes the MLE of the dynamics model from the fixed dataset $\mathcal{D}$, and $\xi$ is a positive constant representing the permissible degree of deviation for the model. Such a constraint set $\mathcal{M}_{\mathcal{D}}$, as defined in Eq. \eqref{eq: constraint_MDP_original}, contributes to mitigating potentially high statistical errors arising from insufficient coverage of the fixed dataset $\mathcal{D}_E$ and avoiding estimating the uncertainty explicitly. By permitting a controlled deviation from the MLE model, the radius $\xi$ achieves a balance between robustness and adaptability, facilitating improved generalization of the dynamics model to unseen state-action pairs while maintaining consistency with the observed data.



\subsubsection{Theoretical Motivation}
In traditional RL theory, let $d_{T, t}^\pi \in \Delta(\mathcal{S} \times \mathcal{A})$ represent the distribution of $(s_t, a_t)$ at time step $t$ under the policy $\pi$ and the dynamics model $T$, encapsulating how the state-action pairs evolve as influenced by both the policy and the environment. Then $d_T^\pi(s, a):= (1 - \gamma) \sum_{t=0}^{\infty} \gamma^t d_{T, t}^\pi$ is the average state action distribution of the policy $\pi$ under the dynamics $T$, which serves as a measure of occupancy. It is commonly assumed that the dataset satisfies global coverage, meaning that the density ratio-based concentrability coefficient $d_T^\pi(s, a) / \rho(s, a)$ is bounded by some constant $C \in \mathbb{R}^{+}$ for \emph{all} policies $\pi \in \Pi$, where $\rho$ represents the state-action distribution from which the dataset $\mathcal{D}$ is sampled. However, this assumption may not hold in practice, as computing an exploratory policy is inherently challenging for large-scale RL problems like CAV scenarios. To address this, we do not assume that $\rho$ provides global coverage. Instead, we focus on a more practical partial coverage setting described below, where the $d_T^{\pi^\ast}(s, a)/{\rho(s, a)}$ is assumed to be bounded for the globally optimal policy $\pi^{\ast}$.

\begin{definition}[Partial Coverage from \cite{uehara2022pessimistic}]
\label{def: Partial_Coverage}
    Partial coverage \cite{chen2019information} holds if there exists an optimal policy $\pi^\ast$ such that:
    \begin{align}
         \sup_{(s, a)} \frac{d_T^{\pi^\ast}(s, a)}{\rho(s, a)} \leq C,\nonumber
    \end{align}
    where $C$ is a constant.
\end{definition}
This partial coverage assumption is significantly weaker than global coverage and more practical. It allows us to compete against the optimal policy $\pi^\ast$ sufficiently represented in a fixed dataset and achieve near-optimal performance under these more realistic assumptions for MBRL scenarios.
Notably, the concentrability coefficient for a policy $\pi$, which is generally adopted for controlling the distribution shift between the fixed dataset distribution and the occupancy measure induced by $\pi$, is computationally intractable. Therefore, we introduce the following new concentrability coefficient based on the model class $\mathcal{M}$ to facilitate computation.
\begin{definition}[Concentrability coefficient from \cite{uehara2022pessimistic}]
\label{def: Concentrability_coefficient}
For an optimal policy $\pi^\ast$, we define the concentrability coefficient $C_{\pi^\ast}$ as:
    \begin{align}
    C_{\pi^\ast}=\max _{T_\phi \in \mathcal{M}_\mathcal{D}} \frac{\mathbb{E}_{(s, a) \sim d_{T}^{\pi^\ast}}\left[\operatorname{TV}\left(T_\phi(\cdot \mid s, a), T(\cdot \mid s, a)\right)^2\right]}{\mathbb{E}_{(s, a) \sim \rho}\left[\operatorname{TV}\left(T_\phi(\cdot \mid s, a), T(\cdot \mid s, a)\right)^2\right]}.\nonumber
\end{align}
\end{definition}


The theoretical analysis from \cite{uehara2022pessimistic} shows that solving a constrained max-min optimization formulation as shown in Eq. \eqref{eq:p_model_based_maxmin_rl} outputs a policy, which, with high probability, is approximately as good as any policy with a state-action distribution covered by the dataset. This is formally stated in the following lemma.
\begin{lemma}[Eq. (9) from \cite{uehara2022pessimistic}]
\label{th: pac_cppo}
Let $T$ denote the true MDP transition function, and define $\pi^\ast$ as the optimal policy. Meanwhile, let $\tilde{T}_\phi$ be the solution MDP obtained from solving the pessimistic model-based min-max RL problem for the dataset $\mathcal{D}$. Then, with probability at least $1 - \delta$, for any MDP $T_\phi \in \mathcal{M}_{\mathcal{D}}$, the following inequality holds:
\begin{align}
    V_{T, \pi^\ast} - V_{\tilde{T}_\phi,\pi^{\ast}}
    &\leq c_1 (1 - \gamma)^{-2} \sqrt{\Lambda_{(s, a) \sim d_{T}^{\pi^\ast},\max}}\nonumber\\
    &\leq c_1 \, (1 - \gamma)^{-2}  \sqrt{C_{\pi^\ast} \Lambda_{(s, a) \sim \rho,\max}},\nonumber
\end{align}
where
\begin{align}
    \Lambda_{(s, a) \sim \rho,\text{max}} = \sup_{T_\phi \in \mathcal{M}_{\mathcal{D}}} \mathbb{E}_{(s, a) \sim \rho} \, \operatorname{TV} \left( T_\phi(\cdot \mid s, a), T(\cdot \mid s, a) \right)^2.\nonumber
\end{align}
Here, $C_{\pi^\ast}$ is defined according to Definition \ref{def: Concentrability_coefficient}, $\rho$ represents the state-action distribution from which the dataset $\mathcal{D}$ is sampled, and $c_1$ is a scaling constant associated with approximation and
concentration bounds.
\end{lemma}

Lemma \ref{th: pac_cppo} shows that if we find a pessimistic model by solving Eq. \eqref{eq:p_model_based_maxmin_rl} under a constrained set of MDPs in Eq. \eqref{eq: constraint_MDP_original}, the performance gap of that model $\tilde{T}_\phi$ and the true dynamics model $T$ under the optimal policy $\pi^\ast$, which has a state-action distribution covered by the dataset, is bounded.

\subsection{System Model}
\label{sec:systemModel}
We propose a mixed autonomy urban traffic system that includes $I$ CAVs and several human-driven vehicles (HVs), along with several intersections as shown in the traffic environment part of Fig. \ref{fig1}. For each vehicle $i$, following the framework of parallel MDPs, the state $s_{t}^{(i)} \in \mathcal{S}^{(i)} = \left(v_t^{(i)},x_t^{(i)}, y_t^{(i)},v_{t,a}^{(i)},v_{t,b}^{(i)},l_{t,a}^{(i)},l_{t,b}^{(i)}, b_t^{(i)} \right)$ at time step $t$ includes the vehicle's velocity $v_t^{(i)} \in \mathbb{R}$, position $z_t^{(i)}=\left(x_t^{(i)}, y_t^{(i)}\right) \in \mathbb{R}^2$, information about the speed and distance of the vehicles ahead and behind it, namely $v_{t,a}^{(i)},v_{t,b}^{(i)},l_{t,a}^{(i)},l_{t,b}^{(i)} \in \mathbb{R}$ and $b_t^{(i)}$ is a boolean variable indicating whether there is a vehicle within the emergency braking range. Each vehicle $i$ is controlled through an adjustable target velocity $v_{t,d}^{(i)} \in \mathbb{R}$ as well as a binary lane change decision $c_t^{(i)}$, denoted as $a_{t}^{(i)}=\left(v_{t,d}^{(i)}, c_t^{(i)}\right)$. Additionally, we assume that both the transition probability and reward functions are unknown and need to be learned through neural networks and interactions with the environment.

Within the framework of decentralized multi-agent MBRL, we assume the availability of history state transition dataset for each agent, denoted as $\mathcal{D}_{E,0:n}^{(i)}=\left\{s_t^{(i)}, a_t^{(i)}, r_t^{(i)},s_{t+1}^{(i)}\right\}_{t=0}^n$. Each agent approximates its transition model $\tilde{T}^{(i)}_\phi, \forall i\in[I]$ using DNNs parameterized by $\phi$, based on its own dataset $\mathcal{D}_{E,0:n}^{(i)}$ and exchanged samples from neighboring agents. In particular, within a graph $\mathcal{G}$ consisting of $I$ CAVs, each CAV can exchange its latest $H$-length\footnote{Notably, the applied length could be episode-dependent but is limited by the maximum value $H$, as will be discussed in Section \ref{sec: MA-PMBRL}.} dataset $\left\{(s_t^{(i)}, a_t^{(i)}, r_t^{(i)}, s_{t+1}^{(i)})\right\}_{t_k-H}^{t_k}$ with neighboring CAVs within a communication range $d \in [0, D(\mathcal{G})]$ at the end of an episode $k$, where $D(\mathcal{G})$ represents the diameter of the graph $\mathcal{G}$. When $d = 0$, the multi-agent MBRL problem reduces to $I$ independent single-agent cases. 
Mathematically, the dataset used for dynamics modeling at each CAV $i$ is updated as: 
\begin{align}
\label{eq: dataset_update}
    \mathcal{D}_{E,0:t_k}^{(i)}:=& \mathcal{D}_{E,0:t_k-H-1}^{(i)} \cup \{s_t^{(i)}, a_t^{(i)}, r_t^{(i)}, s_{t+1}^{(i)}\}_{t_k-H}^{t_k} \nonumber\\
    &\cup_{j \in \mathcal{B}_{t_k}^{(i)}} \{ (s_t^{(j)}, a_t^{(j)}, r_t^{(j)}, s_{t+1}^{(j)})\}_{t_k-H}^{t_k},
\end{align}
where $j \in \mathcal{B}_{t_k}^{(i)}$ represents neighboring CAVs that satisfy the Euclidean distance $\text{dis}_{t_k}(i,j) \leq d$ for between CAVs $i$ and $j$ at the last time-step $t_k$ of episode $k$.

The aforementioned multi-agent settings in a communication range $d$-limited environment transform the pessimistic problem in Eq. \eqref{eq:p_model_based_maxmin_rl} as:
\begin{align}
\label{eq:p_model_based_maxmin_rl_Variant}
   &\tilde{\pi}^{(i)}=\underset{\pi \in \Pi}{\arg \max } \min _{T_\phi \in \mathcal{M}_{\mathcal{D}}} \mathbb{E}_{T_\phi, \pi ,s_0 \sim \mu_0}\left[\sum_{t=0}^{\infty} \gamma^t r^{(i)}_t \mid s^{(i)}_0\right],\nonumber\\ 
   s.t.\quad& T_\phi^{(i)}\propto \mathcal{D}_{E, 0:t}^{(i)}, 0\leq\textit{d}\leq D(\mathcal{G}); \nonumber\\
    & s_0^{(i)} \sim \mu_0; s_{t+1}^{(i)},r^{(i)}_t \sim T^{(i)}_\phi(s_t^{(i)},a_t^{(i)}),\forall t\in[H], 
\end{align}
where $\mathbb{E}(\cdot)$ denotes the expectation operator. This paper aims to provide a modal-based computation-tractable implementation to compute Eq. \eqref{eq:p_model_based_maxmin_rl_Variant} and analyze its theoretical bound.

\section{The \texttt{MA-PMBRL} Algorithm}
\label{sec: MA-PMBRL}
\begin{figure*}[!ht]
    \centering
    \includegraphics[width=1\linewidth]{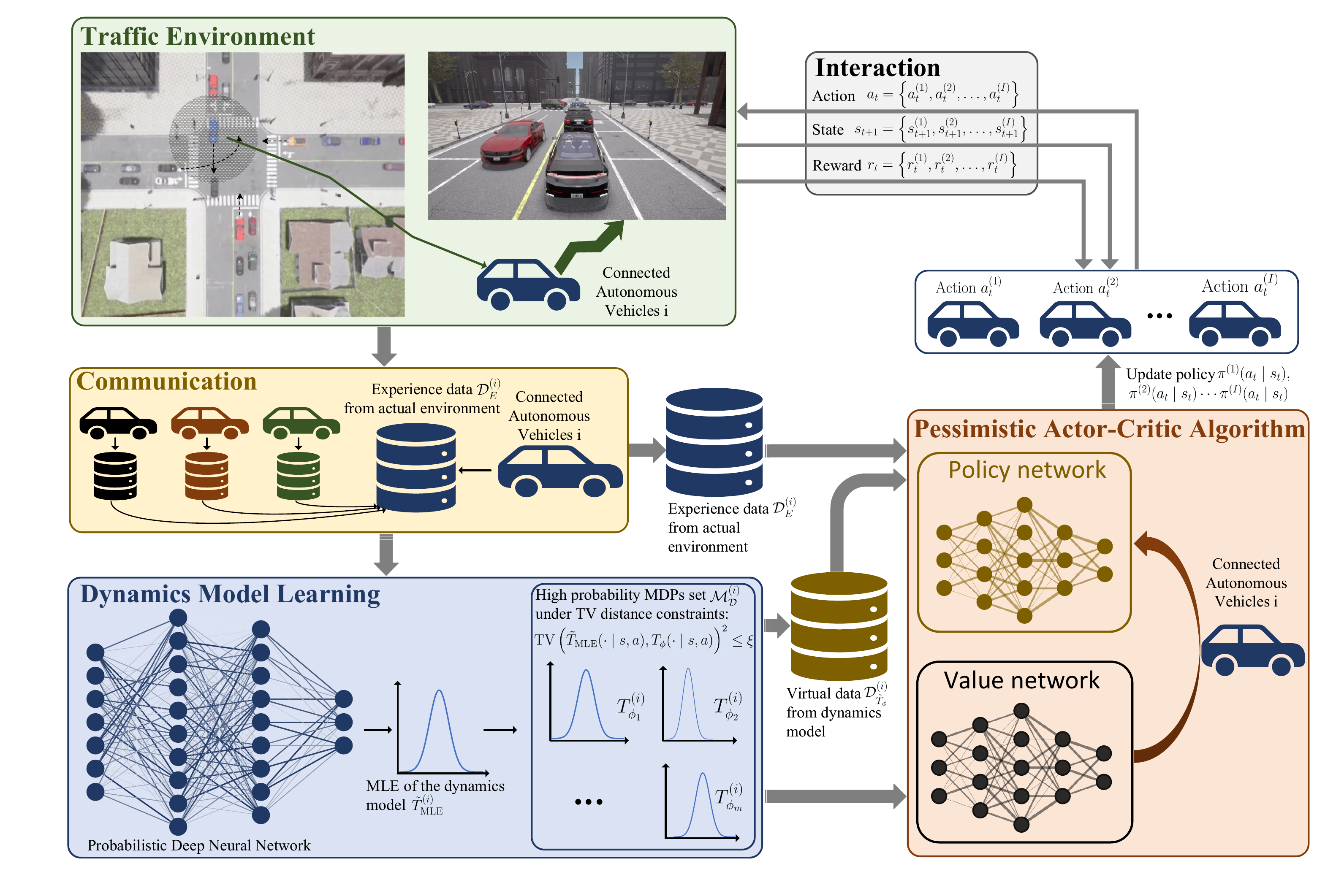}
    \caption{The illustration of the \texttt{MA-PMBRL} algorithm for CAVs.}
    \label{fig1}
\end{figure*}

In this section, we introduce a multi-agent MBRL framework that incorporates the pessimism principle into SAC, tailored for CAV scenarios, where both the dynamics model learning and policy decision-making processes in \texttt{MA-PMBRL} are illustrated in Fig. ~\ref{fig1}.
\subsection{Virtual Environment Dynamics Model Learning}
\label{sec:model_learning}
Within the developed RL framework, an environment model is first constructed to enable rapid, uncertainty-aware prediction of virtual agent-environment interactions. 
Without loss of generality, given available dataset $\mathcal{D}_{E,0:n}^{(i)}=\left\{\left(s_t^{(i)}, a_t^{(i)}, s_{t+1}^{(i)}, r_t^{(i)}\right)\right\}_{t=0}^n$, we approximate the dynamics model at time-step $t$ (i.e., $\tilde{T}_\phi\left(s_{t+1}^{(i)},r_{t}^{(i)}\mid s_t^{(i)},a_t^{(i)}\right),\forall i \in [I]$), using a probabilistic DNN. Notably, assume that the conditional probability distribution of $s_{t+1}^{(i)}$ and $r_{t}^{(i)}$ follows a Gaussian distribution, the Maximum Likelihood Estimate (MLE) $\tilde{T}^{(i)}_{\mathrm{MLE}}\left(\cdot \mid s_t, a_t\right) = \mathcal{N}\left(\tilde{\mu}_\phi\left(s_t, a_t\right), \tilde{\sigma}_\phi^2\left(s_t, a_t\right)\right)$ of the transition model from the dataset can be attained as:
\begin{align}
    \tilde{T}_{\mathrm{MLE}}^{(i)}=\argmin _{(s,a,s',r) \sim \mathcal{D}_{E,0:n}^{(i)}} \sum_{t=1}^n\left(\frac{\left(e_{t}^{(i)}-\tilde{\mu}_\phi \right)^2}{2 \tilde{\sigma}^2_\phi}+ \frac{\ln (2 \pi \tilde{\sigma}^2_\phi)}{2}\right),\nonumber
\end{align}
where $e_{t}^{(i)}$ denote the concatenation the next step's state and reward $(s^{(i)\prime} , r^{(i)})$. 

Next, we compute a constraint set $\mathcal{M}^{(i)}_{\mathcal{D}_E}$, as defined in Eq. \eqref{eq: constraint_MDP_original}. 
Unlike the TV distance, which is less practical to compute and implement in simulations, we use the KL divergence, as it provides a more tractable alternative for measuring the difference between two probability distributions. Thus, Eq. \eqref{eq: constraint_MDP_original} will be reformulated as: 
\begin{align}
\label{eq: constraint_MDP}
    \mathcal{M}_{\mathcal{D}_E}^{(i)}
    =\left\{ T^{(i)}_\phi \mid \mathbb{E}_{(s, a) \sim\mathcal{D}_E^{(i)}}\left[D_{\mathrm{KL}}\left(T^{(i)}_\phi \| \tilde{T}^{(i)}_{\mathrm{MLE}}\right)\right] \leq \xi\right\}
\end{align}
Notably, the KL divergence between a Gaussian transition distribution $T_\phi^{(i)}(\cdot \mid s, a)=\mathcal{N}\left(\mu_\phi\left(s, a\right), \sigma_\phi^2\left(s, a\right)\right)$, 
where $\mu_\phi$ and $\sigma_\phi$ represent the mean and diagonal covariance of the Gaussian distribution respectively, and the MLE estimate $\tilde{T}_{\text {MLE}}(\cdot \mid s, a)$ can be expressed as:
\begin{align}
D_{\mathrm{KL}}\left(T^{(i)}_\phi \| \tilde{T}^{(i)}_{\mathrm{MLE}}\right)
=\frac{\sigma_\phi^2+\left(\mu_\phi-\tilde{\mu}_\phi\right)^2}{2 \tilde{\sigma}_\phi^2}
    +\log \frac{\tilde{\sigma}_\phi}{\sigma_\phi}-\frac{1}{2}.\nonumber
\end{align}

\subsection{Constrained Pessimistic Soft Actor-Critic Algorithm}
\label{sec: al_critic_actor}
Building on the virtual environment model described above, SAC-based RL is the foundational algorithm for the proposed \texttt{MA-PMBRL} to enhance learning efficiency and asymptotic performance.

In \texttt{MA-PMBRL}, the critic is designed to provide a pessimistic evaluation of the value function, drawing on the work of \cite{zanette2021provable}—
At each iteration $t$, given the current policy $\pi_t^{(i)}$, the critic of each agent $i$ selects $\tilde{T}_{\phi}^{(i)} \in \mathcal{M}^{(i)}_{\mathcal{D}_E}$ such that $\tilde{T}_{\phi}^{(i)}=\arg \min _{T_\phi \in \mathcal{M}_{\mathcal{D}}}V_{T_{\phi},\pi_t}^{(i)}$. This introduces pessimism by ensuring that, with high probability, $V_{\tilde{T}_{\phi},\pi_t}^{(i)} \leq V_{T, \pi_t}^{(i)}$, since $T \in \mathcal{M}_{\mathcal{D}^{(i)}_E}$ with high probability. 
Different from \cite{zanette2021provable}, where the critic assumes a \emph{linear} $Q$ function and directly finds a lower bound by minimizing $Q$ within a constrained set of coefficients, we 
minimize a \emph{nonlinear} $Q$ function under an appropriate constraint set through PGD \cite{jain2017non}. Notably, the PGD update of $\phi^{(i)}$ under $\tilde{T}_\phi^{(i)}$\footnote{For simplicity of representation, we slightly abuse the notation $\phi^{(i)}$ for $\tilde{T}_\phi^{(i)}$ and $T_\phi^{(i)}$ here.} after each episode $k$ is performed as follows:
\begin{align}
\label{eq：PGD}
\phi_{x+1}^{(i)}&=\operatorname{Proj}_{\mathcal{M}_{\mathcal{D}_E}^{(i)}}\left(\phi^{(i)\prime}_{x+1}\right)\nonumber\\
&=\operatorname{Proj}_{\mathcal{M}_{\mathcal{D}_E}^{(i)}}\left(\phi_x^{(i)}-\alpha_{\text{PGD}} \nabla_\phi Q_{\tilde{T}_{\phi_x},\pi_t}^{(i)}\right),
\end{align}
where $\operatorname{Proj}_{\mathcal{M}_{\mathcal{D}_E}^{(i)}}\left(\phi^{(i)\prime}_{x+1}\right) = \argmin _{T_\phi \in \mathcal{M}_{\mathcal{D}_E}^{(i)}}\left\|\phi^{(i)\prime}_{x+1}-\phi^{(i)}\right\|_2$ represents the projection of $\tilde{T}_{\phi^{\prime}_{x+1}}^{(i)}$ onto the constraint set $\mathcal{M}_{\mathcal{D}_E}^{(i)}$, ensuring that the updated solution $\phi_{x+1}^{(i)}$ lies within the feasible region defined by $\mathcal{M}_{\mathcal{D}_E}^{(i)}$. 
Our approach can avoid the limitations of Bellman-restricted closeness and extend to more general MDPs with nonlinear $Q$ functions  \cite{uehara2022pessimistic}. Besides, the adoption of PGD, which combines gradient computation and projection to systematically optimize the pessimistic estimation objective while maintaining theoretical rigor and numerical stability, ensures rapid convergence to a feasible solution—after executing a predefined maximum number of PGD updates, denoted as $X$, the algorithm generates a candidate point $\phi_{x}^{(i)}$ at each iteration ($x=1,2,\cdots, X$). Upon completion, the algorithm returns the candidate point that achieves the smallest objective function value\footnote{At the end of $X$ iterations, there are three ways to select the output: return the final iterate $\phi_X$, the average of all iterates $\phi_\mathrm{avg}$, or the point with the smallest objective value $\phi_\mathrm{best}$\cite{jain2017non}. Simulation results in Section \ref{sec: numerical results} demonstrate that selecting the smallest value yields the best performance.}:
\begin{align}
\label{eq: choose_phi}
    \phi^{(i)}_\mathrm{best}= \argmin_{x\in [X]} Q^{(i)}_{\tilde{T}_{\phi_x},\pi_t} 
\end{align}
  
For policy improvement in the actor, the policy function that selects actions to maximize the $Q$-value can be approximated by the policy network $\pi_\psi^{(i)}$. In SAC, the actor-network is typically optimized using the Adam optimizer, which minimizes the gradient of the policy loss $\mathcal{L}_\pi^{(i)}$ concerning its parameters $\psi$, and the optimization objective for the actor policy network is typically achieved by minimizing the following loss function:
\begin{align}
\label{eq: actor_loss}
    \mathcal{L}^{(i)}_\pi =  \mathbb{E}_{s_t \sim   \mathcal{D}_E^{(i)} \cup \mathcal{D}_{\tilde{T}_\phi}^{(i)}, a_t \sim \pi_\psi^{(i)}}& \Bigg[ \alpha_{\text{temp}} \cdot \log \tilde{\pi}^{(i)}\left(a_t \mid s_t\right)\nonumber\\
    &\quad-Q_{\tilde{T}_{\phi},\tilde{\pi}_t}^{(i)}\left(a_t \mid s_t\right) \Bigg]
\end{align}
where $\tilde{T}_\phi^{(i)}$ is a Gaussian distribution derived from the reparameterization trick enabling the policy network to remain differentiable. The term $\tilde{\pi}^{(i)}\left(a_t \mid s_t\right)$ is the output of the policy network (actor), representing the probability of taking action $a_t$ in state $s_t$. $Q_{\tilde{T}_{\phi},\tilde{\pi}_t}^{(i)}$ is the estimated action-value function (critic), and $\alpha_{\text{temp}}$ is the temperature parameter that balances the contribution of policy entropy and the $Q$-value.        

Finally, we summarize our model-based MARL method \texttt{MA-PMBRL} in Algorithm \ref{al: MA-PMBRL}. 
\begin{algorithm}[t]
\caption{The \texttt{MA-PMBRL} algorithm}
\label{al: MA-PMBRL}
\textbf{Initialize:} Communication range $d$, initial state $\mathcal{N}(\mu_{\phi_0},\sigma^2_{\phi_0} )$.\;
\textbf{Initialize:} Vehicle's policy network $\pi_\psi^{(i)}(\cdot \mid s)=\operatorname{Unif}(\mathcal{A}^{(i)})$ and value networks $Q_{\tilde{T}_\phi,\pi}^{(i)}$\;
\textbf{Initialize:} Vehicle's experience buffer $\mathcal{D}_E^{(i)}=\emptyset, \forall i \in[I]$ for storing actual data and $\mathcal{D}_{\tilde{T}_\phi}^{(i)}=\emptyset, \forall i \in[I]$ for storing virtual data\;
Randomly sample action in the initial episode and update $\mathcal{D}_E^{(i)}, \forall i \in [I]$\;

\For{Vehicle $i$ and episode $k$, $\forall i \in [I], \forall k \in [K]$}{
    $\tilde{T}_{\mathrm{MLE}}^{(i)} \leftarrow$ Train the virtual environment MLE dynamics model on $\mathcal{D}_E^{(i)}$ for each vehicle $i$\;
    \For{step $t=1,2, \ldots, H$}{
        \textbf{Generate synthetic $l$-step rollouts:}\\
        Let $l=t+1$\;
        \While{$l \leq t+L_{\text{rollout}}$}{
            Execute  $a_l^{(i)}=\pi^{(i)}_\psi(s_l^{(i)})$ in $ \tilde{T}_{\phi}^{(i)}$\; 
            Get $s_{l+1}^{(i)}, r_l^{(i)}$ and add transition data $(s_l^{(i)}, a_l^{(i)}, r_l^{(i)}, s_{l+1}^{(i)})$ to $\mathcal{D}_{\tilde{T}_\phi}^{(i)}$\;
            Let  $l \leftarrow l+1$\; }
        \textbf{Agent update:} \\
        Select the parameter $\phi_\mathrm{best}^{(i)}$ after performing $X$ iterations of PGD, according to Eq. \eqref{eq：PGD} and Eq. \eqref{eq: choose_phi}\;
        Update $Q_{\tilde{T}_{\phi},\pi}^{(i)}(s, a)$ and $\pi^{(i)}$ using samples from $\mathcal{D}_E^{(i)} \cup \mathcal{D}_{\tilde{T}_\phi}^{(i)}$\;
        Execute  $a_t^{(i)}=\pi^{(i)}_\psi(s_t^{(i)}), \forall i \in [I]$\;
        Obtain reward $r_{t}^{(i)},$ and observe next-step state $s_{t+1}^{(i)}$\;
    }
    Determine neighboring vehicle $j, \forall \text{dis}(i,j) \leq d$ (i.e., within the communication range $d$ of vehicle $i$) to exchange information\;
        Update dataset $\mathcal{D}_E^{(i)}$ by using Eq. \eqref{eq: dataset_update}.\
}
\end{algorithm}

\section{Theoretical Guarantees of \texttt{MA-PMBRL}}
\label{sec: convergence proof}
In this section, we first provide a multi-agent theoretical guarantee of \texttt{MA-PMBRL}, by establishing the suboptimality of the learned policy $\pi_t$ from Algorithm \ref{al: MA-PMBRL} relative to an ideal optimal policy $\pi^{\ast}$, which need not be directly implemented but serves as an upper performance bound or convergence assurance based on the fixed data.
Afterward, 
we derive the group PAC guarantee enabled by communication within a distance of $d$.
\subsection{Methodology}

Beforehand, we introduce the assumption essential for our theoretical analysis to illustrate sample efficiency better. In our multi-agent CAVs framework, data are simultaneously and independently collected by multiple vehicles, each following its strategy. Therefore, the following assumption of the independence of the dataset concerning the collection of fixed data can be made.
\begin{assumption}[Collection of Fixed Dataset]
    \label{as: Data_generation}
    The dataset of each vehicle $i$, $\mathcal{D}_{E,0:n}^{(i)} = \left\{\left(s_t^{(i)}, a_t^{(i)}, r_t^{(i)}, s_{t+1}^{(i)}\right): t=0, \ldots, n\right\}$ satisfies the i.i.d. $\left(s_t^{(i)}, a_t^{(i)}\right) \sim$ $\rho$ with $s_{t+1}^{(i)} \sim T\left(\cdot \mid s_t, a_t\right)$, where $T$ is the true dynamics function of the unknown environment and $\rho$ denotes the distribution induced by the behavior policy under $T$.
\end{assumption}
To facilitate a clear and tractable theoretical analysis of policy optimization, we introduce the following assumption regarding the policy class.
\begin{assumption}[Log-linear Parametric Policy Class] 
\label{as:log_linear_policy} 
The policy class considered in this work is a log-linear parametric family defined as: 
\begin{align}
    \Pi_{\psi} = \left\{ \pi_{\psi}(a|s) = \frac{\exp(\psi \cdot \varphi_{s, a})}{\sum_{a^\prime \in \mathcal{A}}\exp(\psi \cdot \varphi_{s, a^\prime})} \mid \psi \in \mathbb{R}^d \right\}, \nonumber
\end{align}
where the characteristic vector  $\varphi: S\times A \rightarrow \mathbb{R}^d$ satisfies $\|\varphi_{s, a}\|_2\leq \varphi_{\max}$ for any $(s, a)$. This assumption enables tractable theoretical analysis of policy optimization performance. More general parametric policy classes may be adopted under additional assumptions.
\end{assumption}

To find a policy for which we can provide a non-asymptotic bound on its sub-optimality $V_{T, \pi^{\ast}}-\max_{0 \leq t \leq H} V_{T,\pi_t}$, at the end of each episode $k \in [K]$ based on the fixed dataset, we define the group bound as:
\begin{equation}
    \label{eq:group_bound}
    \Delta_{G}^k(H):=\sum_{i=1}^I \left[V^{(i)}_{T, \pi^{\ast}}-\max _{0 \leq t \leq H} V^{(i)}_{T,\pi_t}\right].
\end{equation}
Next, based on the fact that the mean lower bounds maximum, we decompose the cumulative sub-optimality $\Delta_{G}^k(H)$ in Eq. \eqref{eq:group_bound} into the sum of three parts as Eq. \eqref{eq: decompose_gap}. We first investigate each part at the level of an individual agent and then present the group PAC bound. 

\subsection{PAC Bound of Each Agent}
Prominently, Appendix \ref{sec:bound_delta_a_c} states that, consistent with \cite{uehara2022pessimistic}, $\Delta_a^{(i)} \leq c_1 (1-\gamma)^{-2} \sqrt{\frac{C_{\pi^{\ast}} \ln \left( c_2 \left|\mathcal{M}_{\mathcal{D}}\right| / \delta \right)}{|\mathcal{D}_E^{(i)}|}}$ and $\Delta_c^{(i)} \leq 0$. However, bounding $\Delta_b^{(i)}$, which quantifies the difference between the policies $\pi_t^{(i)}$ and the optimal policy $\pi^\ast$ under the same transition model $\tilde{T}_\phi^{(i)} = \argmin_{T_\phi \in \mathcal{M}_{\mathcal{D}}} V_{T_\phi, \pi_t}^{(i)}$, makes a significant difference since we analyze an implementable SAC-based algorithm for gradient optimization while \cite{uehara2022pessimistic} provides only a conceptual framework for pessimistic MBRL without practical implementation aspects and exclusively emphasizes pessimistic model errors (i.e. $\Delta_a^{(i)}$). Additionally, our work provides a detailed comparative analysis with \cite{hongpessimistic}, addressing and improving upon certain limitations in its theoretical derivations. Specifically, we rigorously clarify issues related to the approximation errors arising from the gap between the function class of $Q$ and the linear function class, thereby strengthening and extending the theoretical foundations established in \cite{hongpessimistic}. Therefore, the discrepancy arises mainly from the introduced optimization error during the actor update via gradient descent and the transition error from the model learning process.
\begin{figure*}[t]
\begin{align}
\label{eq: decompose_gap}
    \Delta_{G}^k(H) & :=\sum_{i=1}^I \left[V^{(i)}_{T, \pi^{\ast}}-\max _{0 \leq t \leq H} V^{(i)}_{T,\pi_t}\right]
    \leq \sum_{i=1}^I \left[ \frac{1}{H+1} \sum_{t=0}^H \left( V^{(i)}_{T,\pi^\ast} - V^{(i)}_{T,\pi_t} \right) \right] \nonumber \\
    & = \sum_{i=1}^I \left[ \frac{1}{H+1} \sum_{t=0}^H \Bigg( V^{(i)}_{T,\pi^{\ast}} - \min_{T_\phi \in \mathcal{M}_{\mathcal{D}}} V_{T_\phi,\pi^{\ast}}^{(i)} + \min_{T_\phi \in \mathcal{M}_{\mathcal{D}}} V_{T_\phi,\pi^{\ast}}^{(i)}  - \min_{T_\phi \in \mathcal{M}_{\mathcal{D}}} V_{T_\phi,\pi_t}^{(i)} + \min_{T_\phi \in \mathcal{M}_{\mathcal{D}}} V_{T_\phi,\pi_t}^{(i)} - V_{T,\pi_t}^{(i)} \Bigg) \right] \nonumber \\
    &= \sum_{i=1}^I \Bigg[ \underbrace{\frac{1}{H+1} \sum_{t=0}^H \left( V_{T,\pi^{\ast}}^{(i)} - \min_{T_\phi \in \mathcal{M}_{\mathcal{D}}} V_{T_\phi,\pi^{\ast}}^{(i)} \right)}_{\Delta_a^{(i)}} + \underbrace{\frac{1}{H+1} \sum_{t=0}^H \left( \min_{T_\phi \in \mathcal{M}_{\mathcal{D}}} V_{T_\phi,\pi^{\ast}}^{(i)} - \min_{T_\phi \in \mathcal{M}_{\mathcal{D}}} V_{T_\phi,\pi_t}^{(i)} \right)}_{\Delta_b^{(i)}} \\
    & \quad \quad+ \underbrace{\frac{1}{H+1} \sum_{t=0}^H \left( \min_{T_\phi \in \mathcal{M}_{\mathcal{D}}} V_{T_\phi,\pi_t}^{(i)} - V_{T,\pi_t}^{(i)} \right)}_{\Delta_c^{(i)}} \Bigg] \nonumber 
\end{align}
\hrulefill
\end{figure*}

Based on Lemma \ref{lem: Policy_improvement}, under the umbrella of SAC in Eq. \eqref{eq: actor_loss}, we decompose $\Delta_b^{(i)}$ into three terms (i.e., $\Gamma_a^{(i)}$, $\Gamma_b^{(i)}$, and $\Gamma_c^{(i)}$) as in Eq. \eqref{eq: decompose_delta_b}, for better interpretability and to isolate the components that contribute to the performance gap in the policy. 
\begin{figure*}[t]
\begin{align}
\label{eq: decompose_delta_b}
    \Delta_b^{(i)}&=\frac{1}{H+1} \sum_{t=0}^H \left( \min_{T_\phi \in \mathcal{M}_{\mathcal{D}}} V_{T_\phi,\pi^{\ast}}^{(i)} - \min_{T_\phi \in \mathcal{M}_{\mathcal{D}}} V_{T_\phi,\pi_t}^{(i)} \right)
     =\frac{1}{(H+1)(1-\gamma)} \sum_{t=0}^H \mathbb{E}_{(s, a) \sim d_{\tilde{T}_\phi}^{\pi^\ast}} \left(Q_{\tilde{T}_\phi,\pi_t}^{(i)}(s, a) - V_{\tilde{T}_\phi,\pi_t}^{(i)}(s)\right) \nonumber\\
     &= \frac{1}{(H+1)(1-\gamma)} \sum_{t=0}^H \Bigg[\underbrace{\mathbb{E}_{(s, a) \sim d_{\tilde{T}_\phi}^{\pi^\ast}} \left[Q_{\tilde{T}_\phi,\pi_t}^{(i)}(s, a) - V_{\tilde{T}_\phi,\pi_t}^{(i)}(s)-w_t \cdot \nabla_\psi \log \pi_t^{(i)}(a \mid s)\right]}_{\Gamma_a^{(i)}}\nonumber \\
    &+\underbrace{\mathbb{E}_{(s, a) \sim d_{\tilde{T}_\phi}^{\pi^\ast}}\left[w_t \cdot \nabla_\psi \log \pi_t^{(i)}(a \mid s)\right]-\mathbb{E}_{(s, a) \sim d_{T_\phi}^{\pi^\ast}}\left[w_t \cdot \nabla_\psi \log \pi_t^{(i)}(a \mid s)\right]}_{\Gamma_b^{(i)}} + \underbrace{\mathbb{E}_{(s, a) \sim d_{T_\phi}^{\pi^\ast}}\left[w_t \cdot \nabla_\psi \log \pi_t^{(i)}(a \mid s)\right]}_{\Gamma_c^{(i)}}\Bigg] 
\end{align}
\hrulefill
\end{figure*}
\subsubsection{Bounding $\Gamma_a^{(i)}$}
To estimate an upper bound for $\Gamma_a^{(i)}$, we express it as the difference between two expectations. The first term, which involves the advantage function $Q_{\tilde{T}_\phi, \pi_t}^{(i)}(s, a) - V_{\tilde{T}_\phi, \pi_t}^{(i)}(s)$, is bounded by controlling the variance of policy transitions, ensuring a predictable difference between $Q$ and $V$. The second term, based on Assumption \ref{as:log_linear_policy} and Lemma \ref{lem: log_linear_1}, can be simplified as $w_t \cdot (\varphi_{s, a} - \mathbb{E}_{a^\prime \sim \pi_t(\cdot | s)}[\varphi_{s, a^\prime}])$, represents the policy gradient weighted by $\left\|w_t\right\|_2\leq W$ and the characteristic vector $\left\|\varphi_{s, a}\right\|_2\leq \varphi_{\max}$ for any $(s,a)$. By applying Hölder’s inequality and Lemma \ref{lem: log_linear_1} to the feature difference and approximating the expectation over $(s, a) \sim d_{\tilde{T}_\phi}^{\pi^{\ast}}$ as an average over all actions, we reformulate $\Gamma_a^{(i)}$ as:
\begin{align}
    &\Gamma_a^{(i)}=\mathbb{E}_{(s, a) \sim d_{\tilde{T}_\phi}^{\pi^\ast}} \Bigg[ Q_{\tilde{T}_\phi,\pi_t}^{(i)}(s, a)-\mathbb{E}_{a^{\prime} \sim \pi_t} Q_{\tilde{T}_\phi,\pi_t}^{(i)}\left(s, a^{\prime}\right) \nonumber\\
    &\quad-w_t \cdot \left(\varphi_{s, a}-\mathbb{E}_{a^{\prime} \sim \pi_t(\cdot \mid s)} [\varphi_{s, a^{\prime}}]\right)\Bigg]\nonumber\\
    &\leq \mathbb{E}_{s \sim d_{\tilde{T}_\phi}^{\pi^\ast},a \sim \pi^\ast}  \left[\left| Q_{\tilde{T}_\phi,\pi_t}^{(i)}(s, a)-w_t \cdot\varphi_{s, a}\right| \right]\nonumber\\
    &\quad+\mathbb{E}_{s \sim d_{\tilde{T}_\phi}^{\pi^\ast},a^{\prime} \sim \pi_t}  \left[ \left|w_t \cdot \varphi_{s, a^{\prime}} -Q_{\tilde{T}_\phi,\pi_t}^{(i)}(s, a^{\prime})\right|\right]\nonumber\\
    &\leq 2|\mathcal{A}| \mathbb{E}_{s \sim d_{\tilde{T}_\phi}^{\pi^\ast}, a \sim \operatorname{Unif}(\mathcal{A})}\left[\left|Q_{\tilde{T}_\phi,\pi_t}^{(i)}(s, a)-w_t \cdot \varphi_{s, a}\right|\right]\nonumber\\
    &= 2|\mathcal{A}|\mathbb{E}_{a \sim \operatorname{Unif}(\mathcal{A})}\Bigg[\left( \mathbb{E}_{s \sim d_{\tilde{T}_\phi}^{\pi^\ast}}f^{(i)}(s, a)-\mathbb{E}_{s \sim d_{T}^{\pi^\ast}}{f^{(i)}(s, a)}\right)\nonumber\\
    &\quad+\mathbb{E}_{s \sim d_{T}^{\pi^\ast}}f^{(i)}(s, a)\Bigg],\nonumber
\end{align}
where $f^{(i)}(s, a):=\left|Q_{\tilde{T}_\phi,\pi_t}^{(i)}(s, a)-w_t \cdot \varphi_{s, a}\right|$ be the absolute error term. 

We analyze each component to derive an upper bound for $\Gamma_a^{(i)}$. Following the proof strategy outlined in \cite{hongpessimistic} and leveraging Lemma \ref{lem: generalized_simulation}, we present the detailed derivation as follows:
\begin{itemize}
\item 
We note that as $a \sim \operatorname{Unif}(\mathcal{A})$, $
    \mathbb{E}_{s \sim d_{\tilde{T}_\phi}^{\pi^\ast},a \sim \operatorname{Unif}(\mathcal{A})}f^{(i)}(s, a)=\mathbb{E}_{s \sim d_{\tilde{T}_\phi}^{\pi^\ast}}\left[\frac{1}{|\mathcal{A}|}\sum_a f^{(i)}(s, a)\right]$. 
By defining an induced MDP with a modified reward function $\tilde{r}^{(i)}(s,a):= \tilde{f}^{(i)}(s)= \frac{1}{|\mathcal{A}|}\sum_a f^{(i)}(s, a)$, which depends only on 
$s$ and remains the same for all $a$, we obtain:
\begin{align}
    \mathbb{E}_{s \sim d_{\tilde{T}_\phi}^{\pi^\ast}}\tilde{f}^{(i)}(s)&=\mathbb{E}_{s \sim d_{\tilde{T}_\phi}^{\pi^\ast}}\tilde{r}^{(i)}(s,a)
\nonumber\\
&=\mathbb{E}_{s \sim d_{\tilde{T}_\phi}^{\pi^\ast},a \sim \pi^\ast}\tilde{r}^{(i)}(s,a).\nonumber
\end{align}
We derive an upper bound on the expected value difference by state-action trajectory sampling. Specifically, 
we have the bound $\tilde{f} \leq | W\cdot\|\varphi_{s, a}\|_{2, d_{T\circ \mathrm{Unif}}^\pi}+R_{\max}/(1-\gamma)|$, where $\|\varphi_{s, a}\|_{2, d_{T\circ \mathrm{Unif}}^\pi}$ denotes the $L_2$ -norm of the feature under the distribution $d_{T\circ \mathrm{Unif}}^\pi$. Thus we can establish the following upper bound for the value function induced by the modified reward function $\tilde{r}^{(i)}$ as: 
\begin{align}
    \tilde{V} \leq \frac{1}{1-\gamma}\cdot\left| W\cdot\|\varphi_{s, a}\|_{2, d_{T\circ \mathrm{Unif}}^\pi}+\frac{R_{\max}}{1-\gamma}\right|.\nonumber
\end{align}
Hence, based on Definition \ref{def: Concentrability_coefficient} and Lemma \ref{lem: generalized_simulation}, we obtain: 
\begin{align}
\label{eq: Gamma_a_p1}
    &\mathbb{E}_{\substack{s \sim d_{\tilde{T}_\phi}^{\pi^\ast}\\ a \sim \operatorname{Unif}(\mathcal{A})}}f^{(i)}(s, a)-\mathbb{E}_{\substack{s \sim d_{T}^{\pi^\ast}\\a \sim \operatorname{Unif}(\mathcal{A})}}{f^{(i)}(s, a)}  \nonumber\\
    & =\mathbb{E}_{s \sim d_{\tilde{T}_\phi}^{\pi^\ast}}\frac{1}{|\mathcal{A}|}\sum_a f^{(i)}(s, a)-\mathbb{E}_{s \sim d_{T}^{\pi^\ast}}\frac{1}{|\mathcal{A}|}\sum_a f^{(i)}(s, a)\nonumber\\
    &\leq \frac{2\gamma}{(1-\gamma)^2}\left[W\cdot\|\varphi_{s, a}\|_{2, d_{T\circ \mathrm{Unif}}^\pi}+\frac{R_{\max}}{1-\gamma}\right]\nonumber\\
    &\qquad \cdot \sqrt{C_{\pi^\ast}}\mathbb{E}_{(s, a) \sim \rho}\left[\operatorname{TV}\left(\tilde{T}_\phi(\cdot \mid s, a), T(\cdot \mid s, a)\right)^2\right]\nonumber\\
    &\leq \frac{2\gamma}{(1-\gamma)^2}\left[W\cdot\|\varphi_{s, a}\|_{2, d_{T\circ \mathrm{Unif}}^\pi}+\frac{R_{\max}}{1-\gamma}\right]\nonumber\\
    &\qquad \cdot \sqrt{\frac{C_{\pi^{\ast}} \ln \left( c_2 \left|\mathcal{M}_{\mathcal{D}}\right| / \delta \right)}{|\mathcal{D}_E^{(i)}|}}\\
    & \leq\Bigg[\frac{2\gamma c_3}{(1-\gamma)^2}+\frac{2\gamma R_{\max}}{(1-\gamma)^3}\Bigg]\cdot \sqrt{\frac{C_{\pi^{\ast}} \ln \left( c_2 \left|\mathcal{M}_{\mathcal{D}}\right| / \delta \right)}{|\mathcal{D}_E^{(i)}|}}.\nonumber
\end{align}
where $W$ is the upper bound on the weight norm, $|\mathcal{D}_E^{(i)}|$ is the sample size of the dataset $\mathcal{D}_E^{(i)}$, and $c_3= W\cdot\varphi_{\max}$.
\item To analyze the second part of $\Gamma_a^{(i)}$, we establish a relationship between the distributions induced by the two policies by expressing it through their density ratio. 
Applying Lemma \ref{lem: Distribution_Conversion_Lemma}, we bound the expectation over the uniform action distribution and the stationary distribution of $\pi^{\ast}$. Using the triangle inequality, we obtain
\begin{align}
\label{eq: Gamma_a_p2}
    &\mathbb{E}_{s \sim d_{T}^{\pi^{\ast}}, a \sim \operatorname{Unif}(\mathcal{A})}[f^{(i)}(s, a)] \nonumber\\
    &\leq  \left\|\frac{d_{T}^{\pi^\ast}(s)}{d_{\tilde{T}_\phi}^{\pi_t}(s)}\right\|_{\infty}^{\frac{1}{2}}
    \sqrt{\mathbb{E}_{s \sim d_{\tilde{T}_\phi}^{\pi_t}, a \sim \operatorname{Unif}(\mathcal{A})}\left[f^{(i)}(s, a)^2\right]}\nonumber\\
    & \leq \left\|\frac{d_{T}^{\pi^\ast}}{d_{\tilde{T}_\phi}^{\pi_t}}\right\|_{\infty}^{\frac{1}{2}} \sqrt{2}\sqrt{\mathbb{E}_{s, a}\left[\left|Q^{(i)}_{\tilde{T}_\phi,\pi_t}(s, a)\right|^2\right]+c_3^2}\nonumber\\
    & \leq \left\|\frac{2d_{T}^{\pi^\ast}}{d_{\tilde{T}_\phi}^{\pi_t}}\right\|_{\infty}^{\frac{1}{2}}\bigg[\sqrt{\mathbb{E}_{s, a}\left[\left|Q^{(i)}_{\tilde{T}_\phi,\pi_t}(s, a)\right|^2\right]}+c_3\bigg],
\end{align}
where the last inequality comes directly from applying the generalized triangle inequality. 
Subsequently, we apply Hoeffding’s inequality together with Lemma \ref{lem: Boundedness_Q} to quantify the generalization error as follows:
\begin{align}
\label{eq: bound_empirical_Q}
    &\sqrt{\mathbb{E}_{s, a}\left[\left|Q^{(i)}_{\tilde{T}_\phi,\pi_t}(s, a)\right|^2\right]}\nonumber\\
    & \leq \sqrt{ \sum_{(s, a) \in \mathcal{D}_E^{(i)}}\frac{|Q^{(i)}(s, a)|^2}{\left|\mathcal{D}_E^{(i)}\right|}}+\frac{R_{\max}}{1-\gamma}\left(\frac{\ln (2 / \delta)}{2\left|\mathcal{D}_E^{(i)}\right|}\right)^{1 / 4}\nonumber\\
    &\leq \left[\frac{\left|\mathcal{D}_E^{(i)}\right| \cdot \left(\frac{R_{\max}}{1-\gamma}\right)^2}{\left|\mathcal{D}_E^{(i)}\right|}\right]^{1/2}+\frac{R_{\max}}{1-\gamma}\left(\frac{\ln (2 / \delta)}{2\left|\mathcal{D}_E^{(i)}\right|}\right)^{1 / 4}\nonumber\\
    & = \frac{R_{\max}}{1-\gamma} + \frac{R_{\max}}{1-\gamma}\left(\frac{\ln (2 / \delta)}{2\left|\mathcal{D}_E^{(i)}\right|}\right)^{1 / 4}.
\end{align}

By merging Eq. \eqref{eq: Gamma_a_p2} with Eq. \eqref{eq: bound_empirical_Q}, we obtain an upper bound for the second term of $\Gamma_a^{(i)}$ as shown in Eq. \eqref{eq: Gamma_a_p2_final}. Unlike \cite{hongpessimistic}, our approach leverages empirical estimation and a rigorous generalization error analysis based on Hoeffding's inequality to characterize the gap between the function class of $Q(s, a)$ and the linear function class $w_t \cdot \varphi_{s, a}$. With high probability, we have:
\begin{align}
\label{eq: Gamma_a_p2_final}
&\mathbb{E}_{s \sim d_{T}^{\pi^{\ast}}, a \sim \operatorname{Unif}(\mathcal{A})}[f^{(i)}(s, a)] \\
\lesssim& \frac{\sqrt{2}}{1-\gamma}\left\|\frac{d_{T}^{\pi^\ast}}{\mu_0}\right\|_{\infty}^{\frac{1}{2}}\Bigg[\frac{R_{\max}}{1-\gamma}\left(1+\left(\frac{\ln (2 / \delta)}{2\left|\mathcal{D}_E^{(i)}\right|}\right)^{1 / 4}\right)+c_3\Bigg].\nonumber
\end{align}
\end{itemize}
Thus, by combining these results, we can upper-bound the original term by summing up the contributions from the two parts, Eq. \eqref{eq: Gamma_a_p1} and Eq. \eqref{eq: Gamma_a_p2_final}, we have:
\begin{align}
\label{eq: Gamma_a}
   & \Gamma_a^{(i)} \leq \left[\frac{4\gamma c_3|\mathcal{A}|}{(1-\gamma)^2}+\frac{4\gamma|\mathcal{A}|R_{\max}}{(1-\gamma)^3}\right]  \sqrt{\frac{C_{\pi^{\ast}} \ln \left( c_2 \left|\mathcal{M}_{\mathcal{D}}\right| / \delta \right)}{|\mathcal{D}_E^{(i)}|}}\nonumber\\
   &+\frac{2\sqrt{2}|\mathcal{A}|}{1-\gamma}\left\|\frac{d_{T}^{\pi^\ast}}{\mu_0}\right\|_{\infty}^{\frac{1}{2}}\Bigg[\frac{R_{\max}}{1-\gamma} + \frac{R_{\max}}{1-\gamma}\left(\frac{\ln (2 / \delta)}{2\left|\mathcal{D}_E^{(i)}\right|}\right)^{1 / 4}+c_3\Bigg].
\end{align}
\subsubsection{Bounding $\Gamma_b^{(i)}$}
Notably, $\Gamma_b^{(i)}$ measures the difference between two expectations over the visitation measures under transition models $\tilde{T}_\phi^{(i)}$ and $T$, that is,
\begin{align}
    \Gamma_b^{(i)}&=\mathbb{E}_{(s, a) \sim d_{\tilde{T}_\phi}^{\pi^\ast}}\left[w_t \cdot \nabla_\psi \log \pi_t^{(i)}(a\mid s)\right]\nonumber\\
    &\quad-\mathbb{E}_{(s, a) \sim d_{T}^{\pi^\ast}}\left[w_t \cdot \nabla_\psi \log \pi_t^{(i)}(a\mid s)\right].\nonumber
\end{align}

Using trajectories generated by $\pi^\ast$ under the distributions  $\tilde{T}_\phi^{(i)}$ and $T$, respectively, we can introduce the another reward function $\tilde{r}^{(i)}(s, a):= w_t \cdot \nabla_\psi \log \pi_t^{(i)} (a\mid s)$. 
Based on Assumption \ref{as:log_linear_policy} and Lemma \ref{lem: log_linear_1}, we derive the following upper bound for $\tilde{r}^{(i)}(s, a)$ as:
\begin{align}
    \left|\tilde{r}^{(i)}(s, a) \right|&=\left|w_t \cdot \nabla_\psi \log \pi_t^{(i)} (a\mid s) \right|\nonumber\\
    &=\left\|w_t\right\|_2\cdot \left\| \varphi_{s, a}-\mathbb{E}_{a^{\prime} \sim \pi_t^{(i)}(\cdot \mid s)} \varphi_{s, a^{\prime}} \right\|\nonumber\\
    &\leq 2W\cdot \varphi_{\max}=2c_3.\nonumber
\end{align}
Consequently, the value function is bounded by $ \left| \tilde{V}_{T_\phi,\pi_t}^{(i)}(s) \right|\leq \frac{2c_3}{1-\gamma}$ for all $s \in \mathcal{S}$.

Similar to the bound of $\Gamma_a^{(i)}$, we can 
incorporate Lemma \ref{th: pac_cppo} and Lemma \ref{lem: generalized_simulation} to bound $\Gamma_b^{(i)}$ in terms of the total variation distance between $\tilde{T}_\phi^{(i)}$ and $T$, and finally obtain:
\begin{align}
\label{eq: Gamma_b}
&\Gamma_b^{(i)}\nonumber\\
= &\mathbb{E}_{\tau \sim \tilde{T}_\phi^{(i)}}\left[\sum_{t=0}^{H} \gamma^t \tilde{r}^{(i)}\left(s_t, a_t\right)\right]-\mathbb{E}_{\tau \sim T}\left[\sum_{t=0}^{H} \gamma^t \tilde{r}^{(i)}\left(s_t, a_t\right)\right]\nonumber \nonumber\\
=&\left|\tilde{V}_{\tilde{T}_\phi,\pi_t}^{(i)} - \tilde{V}_{T,\pi_t}^{(i)}\right|\nonumber\\
     \leq &\frac{4c_3 \gamma}{(1 - \gamma)^2} \mathbb{E}_{(s, a) \sim d^{\pi_t}_{\tilde{T}_\phi}} \left[ \text{TV}(\tilde{T}_\phi^{(i)}(\cdot | s, a), T(\cdot | s, a)) \right]\nonumber\\
    \leq &4c_3 \gamma (1-\gamma)^{-2} \sqrt{\frac{C_{\pi^{\ast}} \ln \left( c_2 \left|\mathcal{M}_{\mathcal{D}}\right| / \delta \right)}{|\mathcal{D}_E^{(i)}|}}.
\end{align}

\subsubsection{Bounding $\Gamma_c^{(i)}$}
We start from the definition of  $\Gamma_c^{(i)} = \mathbb{E}_{(s,a) \sim d_{T}^{\pi^\ast}} \left[ w_t \cdot \nabla_\psi \log \pi_t^{(i)}(a|s) \right]
$. By Lemma \ref{lem: log_linear_2},  $\log \pi_t^{(i)}(a|s)$ is $\varphi^2_{\max}$-smooth. 
By substituting the gradient update $\psi_{t+1} - \psi_t = \eta w_t$ and applying the smoothness property in Lemma \ref{lem: smoothness_property}, we can have:
\begin{align}
     & \Gamma_c^{(i)} = \frac{1}{\eta}\mathbb{E}_{(s,a) \sim d_{T}^{\pi^\ast}}\left[ \nabla_\psi \log \pi_t^{(i)}(a|s) \cdot (\psi_{t+1} - \psi_t) \right]\nonumber \\
     \leq &\frac{1}{\eta}\mathbb{E}_{(s,a) \sim d_{T}^{\pi^\ast}} \left[ \log \pi_{t+1}^{(i)}(a|s) - \log \pi_{t}^{(i)}(a|s) \right] \nonumber \\
    &\quad +\frac{\varphi_{\max}^2}{2\eta} \|\psi_{t+1} - \psi_t\|^2 \nonumber\\
    \leq& \frac{1}{\eta}\mathbb{E}_{(s,a) \sim d_{T}^{\pi^\ast}} \left[ \log \pi_{t+1}^{(i)}(a|s) - \log \pi_{t}^{(i)}(a|s) \right]  + \frac{c_3^2\eta}{2} \nonumber\\
    = &\frac{1}{\eta}\mathbb{E}_{(s,a) \sim d_{T}^{\pi^\ast}} \Bigg[ \text{D}_\text{KL} \left(\pi^{(i)}_{t+1}(\cdot | s) \| \pi^{\ast}(\cdot | s) \right)\nonumber\\
    &\quad-\text{D}_\text{KL} \left(\pi^{(i)}_{t}(\cdot | s) \| \pi^{\ast}(\cdot | s) \right) \Bigg]  + \frac{c_3^2\eta}{2}.\nonumber
\end{align}

Taking the expectation of $\Gamma_c^{(i)}$  and averaging over $t = 0, \cdots, H$, we arrive at the following bound:
\begin{align}
\label{eq: Gamma_c}
    \frac{1}{H+1} \sum_{t=0}^H \Gamma_c^{(i)} & \leq \frac{\mathbb{E}_{s \sim d_{T}^{\pi^\ast}} \left[\text{D}_\text{KL} \left(\pi_0^{(i)}(\cdot | s) \| \pi^{\ast}(\cdot | s) \right) \right]}{\eta(H+1)} +  \frac{c_3^2\eta }{2}\nonumber\\
    &\leq \frac{c_3^2\eta}{2} + \frac{\log |\mathcal{A}|}{\eta(H+1)},
\end{align}
where we take account of the initial policy $ \pi_0(\cdot|s) = \text{Uniform}(\mathcal{A})$, and correspondingly have $\text{D}_\text{KL} \left( \pi_0 \| \pi^{\ast} \right) \leq \log |\mathcal{A}|$.

By combining Eq. \eqref{eq: Gamma_a}, Eq. \eqref{eq: Gamma_b}, and Eq. \eqref{eq: Gamma_c}, we can derive the following upper bound for $\Delta_b^{(i)}$:
\begin{align}
    &\Delta_b^{(i)}\leq \frac{1}{(H+1)(1-\gamma)} \sum_{t=0}^H\Bigg[\Gamma_a^{(i)}+\Gamma_b^{(i)}+\Gamma_c^{(i)}\Bigg]\nonumber\\
    &\leq \left[\frac{4c_3\gamma\left( |\mathcal{A}|+1\right)}{(1-\gamma)^3}+\frac{4\gamma|\mathcal{A}|R_{\max}}{(1-\gamma)^4}\right]  \sqrt{\frac{C_{\pi^{\ast}} \ln \left( c_2 \left|\mathcal{M}_{\mathcal{D}}\right| / \delta \right)}{|\mathcal{D}_E^{(i)}|}}\nonumber\\
   &+\frac{2\sqrt{2}|\mathcal{A}|}{(1-\gamma)^2}\left\|\frac{d_{T}^{\pi^\ast}}{\mu_0}\right\|_{\infty}^{\frac{1}{2}}\cdot\Bigg[\frac{R_{\max}}{1-\gamma} + \frac{R_{\max}}{1-\gamma}\left(\frac{\ln (2 / \delta)}{2\left|\mathcal{D}_E^{(i)}\right|}\right)^{1 / 4}+c_3\Bigg]\nonumber\\
   &+\frac{c_3^2\eta}{2(1-\gamma)}+ \frac{\log |\mathcal{A}|}{\eta(H+1)(1-\gamma)}.\nonumber
\end{align}
Thus, the summarized upper bounds $\Delta_a^{(i)}, \Delta_b^{(i)}$ and $\Delta_c^{(i)}$  for PAC bound of single agent as the following Lemma.
\begin{lemma}
\label{lem: PAC_bound_single_agent}
Let $\eta$ and $\xi$ be some constant such that $T \in \mathcal{M}_{\mathcal{D}}$ holds with high probability at least $1-\delta$. Suppose that the characteristic vector and policy gradient weights are bounded, i.e., $\left\|\varphi_{s, a}\right\|_2\leq \varphi_{\max}$ and $\left\|w_t\right\|_2\leq W$, and define $c_3= W\cdot\varphi_{\max}$. Then under Assumption \ref{as: Data_generation} and Assumption \ref{as:log_linear_policy}, there exist positive constants $c_1$ and $c_2$ such that:
    \begin{align}
    &V^{(i)}_{T, \pi^{\ast}}-\max _{0 \leq t \leq H} V^{(i)}_{T,\pi_t} \leq \sqrt{\frac{C_{\pi^{\ast}} \ln \left( c_2 \left|\mathcal{M}_{\mathcal{D}}\right| / \delta \right)}{|\mathcal{D}_E^{(i)}|}} \nonumber\\
    &\cdot 
    \left[\frac{c_1}{(1-\gamma)^{2}}+\frac{4c_3\gamma\left( |\mathcal{A}|+1\right)}{(1-\gamma)^3}+\frac{4\gamma|\mathcal{A}|R_{\max}}{(1-\gamma)^4}\right]\nonumber\\
   &+\frac{2\sqrt{2}|\mathcal{A}|}{(1-\gamma)^2}\left\|\frac{d_{T}^{\pi^\ast}}{\mu_0}\right\|_{\infty}^{\frac{1}{2}}\cdot\Bigg[\frac{R_{\max}}{1-\gamma} + \frac{R_{\max}}{1-\gamma}\left(\frac{\ln (2 / \delta)}{2\left|\mathcal{D}_E^{(i)}\right|}\right)^{1 / 4}+c_3\Bigg]\nonumber\\
   &+\frac{c_3^2\eta}{2(1-\gamma)}+ \frac{\log |\mathcal{A}|}{\eta(H+1)(1-\gamma)}.\nonumber
\end{align}
\end{lemma}

\subsection{Group PAC Bounds}
We assume all CAVs with constrained communication range can form a graph structure. Specifically, for each episode $k$, 
a neighborhood graph, denoted as $\mathcal{G}_{d,k}^{(i)}=\left(\mathcal{F}_{d,k}^{(i)},\mathcal{E}^{(i)}_{d,k}\right)\subset\mathcal{G} $, is constructed. Here, $\mathcal{E}^{(i)}_{d,k}$ is the set of communication links between agent $i$ and other agents $i^{\prime} \in \mathcal{F}_{d,k}^{(i)}$. We define the total number of samples $N_{d,k}^{(i)}$ for vehicle $i$ after communication as the cumulative sum of observations collected from itself and its neighbors within the communication range $d\in [0, D(\mathcal{G}_k)]$. Therefore, at the end of episode $k$,  
\begin{align}
    N_{d,k}^{(i)}:=N_{d,k-1}^{(i)} +\sum\nolimits_{i^{\prime} \in \mathcal{G}_{d,k}^{(i)}} |\mathcal{D}_{E, k}^{(i^{\prime})}|,\nonumber
\end{align}
where $\mathcal{D}_{E, k}^{(i^{\prime})}$ represents the data collected by vehicle $i$ during its interactions with the environment in episode $k$.

To address the challenge of directly estimating $N_{d,k}^{(i)}$, we introduce the concept of a clique cover \cite{CORNEIL1988109} for $\mathcal{F}_{d,k}^{(i)}$, which is a collection of cliques that collectively cover all vertices of a power graph. Let $\mathbf{C}_{d,k}$ denote a clique cover of $\mathcal{G}_{d,k}$, where the graph contains multiple maximal cliques $\mathcal{C}_{d,k}^{1}, \mathcal{C}_{d,k}^{2}, \cdots, \mathcal{C}_{d,k}^{I^\prime} \in \mathbf{C}_{d,k}$ with $I^\prime\leq I$. The size of each clique $\mathcal{C}_{d,k}^{i^\prime}$ is denoted by $|\mathcal{C}_{d,k}^{i^\prime}|$. For each clique $\mathcal{C}_{d,k}^{i^\prime}$, let $N_{\mathcal{C},k}^{i^\prime}$ represent the total number of samples exchanged within the clique $\mathcal{C}_{d,k}^{i^\prime}$ at the end of the episode $k$, which are accessible to all vehicles $i \in \mathcal{C}_{d,k}^{i^\prime}$. To simplify the computation, we approximate $N_{\mathcal{C},k}^{i^\prime}\approx |\mathcal{C}_{d,k}^{i^\prime}|kH$, which satisfies the definition. The clique cover number $\bar{\chi}\left(\mathcal{G}_{d,k}\right)$ is defined as the minimum number of cliques required to cover the power graph $\mathcal{G}_{d,k}$ under the communication range. Furthermore, we define $\bar{\chi}\left(\mathcal{G}_d\right):= \max_{k \in [K]} \{\bar{\chi}\left(\mathcal{G}_{d,k}\right)\}$ as the maximum clique cover number across all episodes.

Thus, by utilizing the definition of cliques and bounding the contributions within each clique according to Lemma \ref{lemma:clique_number}, we can have
\begin{align}
    \sum_{i=1}^I\frac{1}{\sqrt{|\mathcal{D}_E^{(i)}|}}&= \sum_{i^\prime=1}^{I^\prime} \sum_{i \in \mathcal{C}_{d,k}^{i^\prime}}\frac{1}{\sqrt{N_{d,k}^{(i)}}}\nonumber\\
    &\leq \sum_{i^\prime=1}^{I^\prime} |\mathcal{C}_{d,k}^{i^\prime}| \frac{1}{\sqrt{N_{\mathcal{C},k}^{i^\prime}}}\lesssim\sqrt{\bar{\chi}\left(\mathcal{G}_d\right)I/{kH}}.\nonumber
\end{align}

We now present Theorem \ref{Th: pac_multi_agent}, which provides a theoretical guarantee for the multi-agent system, establishing an upper bound on the sub-optimality of the optimal policy among the iterated policies in \texttt{MA-PMBRL} \ref{al: MA-PMBRL} at the end of each episode.

\begin{theorem}[PAC bound for \texttt{MA-PMBRL}]
\label{Th: pac_multi_agent}
Let $\eta$ and $\xi$ be constants chosen such that $T \in \mathcal{M}_{\mathcal{D}}$ holds with probability at least $1-\delta$. Assume the policy gradient satisfies $\left\|w_t\right\|_2\leq W$, and the feature vector satisfies $\left\|\varphi_{s, a}\right\|_2 \leq \varphi_{\max}$. Define constants $c_1, c_2$ and $ c_3= W\cdot \varphi_{\max}$. Then, at the end of the \(k\)-th episode with a maximum length \(H\), the algorithm's performance gap:
\begin{align}
     \Delta_{G}^k(H)&=\sum_{i=1}^I \left[ \frac{1}{H+1} \sum_{t=0}^H \left( V^{(i)}_{T,\pi^{\ast}} - V^{(i)}_{T,\pi_t} \right) \right],\nonumber
\end{align}
satisfies the following bound:
\begin{align}
\label{eq: group_pac_bound}
\Delta_{G}^k(H)& \leq\left[\frac{c_1}{(1-\gamma)^{2}}+\frac{4c_3\gamma\left( |\mathcal{A}|+1\right)}{(1-\gamma)^3}+\frac{4\gamma|\mathcal{A}|R_{\max}}{(1-\gamma)^4}\right]\nonumber\\
&\cdot\sqrt{{C_{\pi^{\ast}}\bar{\chi}\left(\mathcal{G}_d\right)I \ln \left( c_2 \left|\mathcal{M}_{\mathcal{D}}\right| / \delta \right)}/{kH}}\nonumber\\
&+\frac{2\sqrt{2}|\mathcal{A}|}{(1-\gamma)^2}\left\|\frac{d_{T}^{\pi^\ast}}{\mu_0}\right\|_{\infty}^{\frac{1}{2}}\Bigg[ \frac{R_{\max}}{1-\gamma}\sqrt[4]{\frac{\ln (\frac{2}{\delta})\bar{\chi}\left(\mathcal{G}_d\right)I}{2kH}}\\
&+\frac{IR_{\max}}{1-\gamma} +c_3I\Bigg]+\frac{c_3^2\eta I }{2(1-\gamma)}+ \frac{I\log |\mathcal{A}|}{\eta(H+1)(1-\gamma)}.\nonumber
\end{align}
\end{theorem}

In summary, we conclude the proof of Theorem \ref{Th: pac_multi_agent}, establishing a group PAC guarantee bound in Eq. \eqref{eq: group_pac_bound} for general parallel MDPs under the partial coverage assumption $C_{\pi^{\ast}}\leq \infty$, relying solely on a realizable hypothesis class, with probability at least $1-\delta$. Theorem \ref{Th: pac_multi_agent} explicitly characterizes the evolution of the performance gap $\Delta_{G}^k(H)$ both over episode length and across agents. In contrast, within traditional non-communicative reinforcement learning frameworks, the group PAC bound typically accumulates additively from individual single-agent bounds, highlighting the substantial benefit of incorporating communication among agents in reducing overall complexity. Moreover, the explicit inclusion of the clique cover term $\bar{\chi}\left(\mathcal{G}_d\right)$ highlights how structured communication among agents effectively reduces sample complexity, underscoring the strategic benefit of exploiting inter-agent dependencies to enhance policy optimization and accelerate convergence in MARL.

\section{Experimental Settings and Numerical Results}
\label{sec: numerical experiments}
\subsection{Experimental Settings}

\begin{figure}
	\centering
	\subfigure[Aerial View]{
		\begin{minipage}[t]{0.4\linewidth}
			\centering
			\includegraphics[height=1.2in]{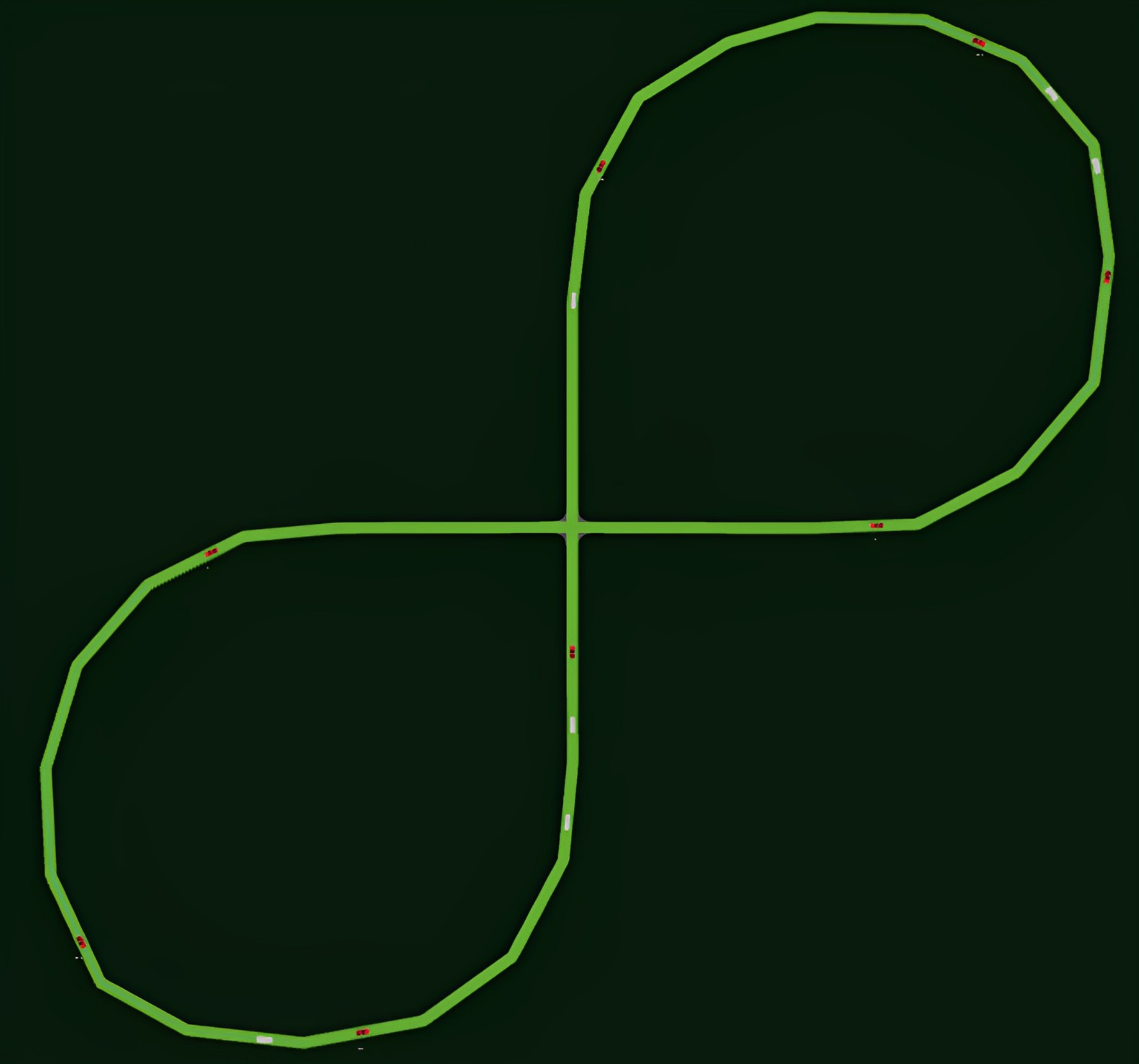}
		\end{minipage}
	}%
	\subfigure[Regional Enlarged View]{
		\begin{minipage}[t]{0.45\linewidth}
			\centering
			\includegraphics[height=1.2in]{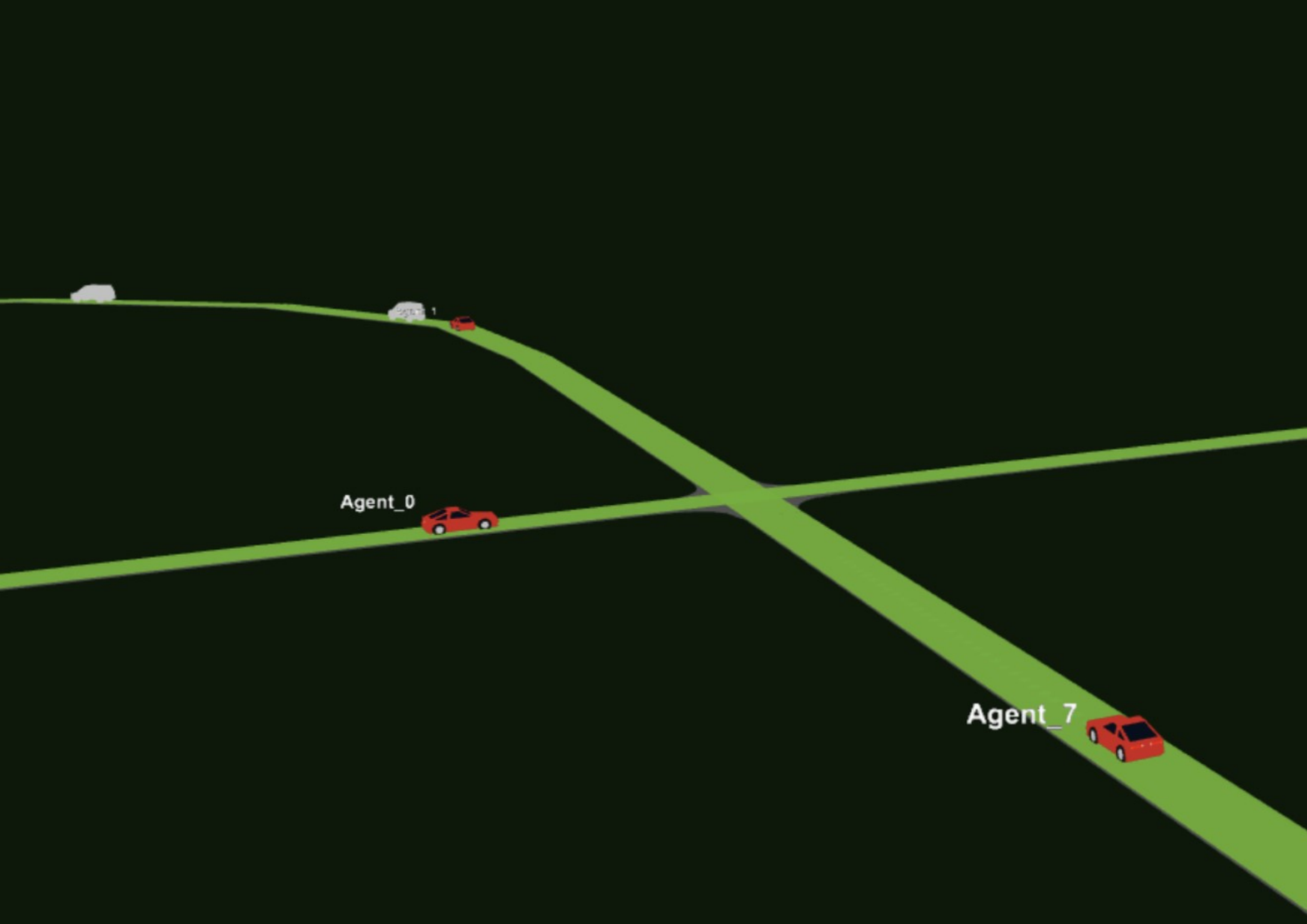}
		\end{minipage}
	}%
	\centering
	\caption{The ``Unprotected Intersection'' scenario in the closed ``Figure Eight" loop for Simulations. (a) presents an aerial view of the ``Figure Eight" loop, while (b) provides the regional enlarged view of the ``Unprotected Intersection''.}
 \label{fig3}
\end{figure}

In this section, we present comprehensive performance evaluations of our proposed \texttt{MA-PMBRL} framework within autonomous vehicle control applications, systematically comparing its effectiveness against leading RL approaches including \texttt{MA-PETS}~\cite{wen_multiagent_2024}, \texttt{MBPO}~\cite{MBPO}, \texttt{FIRL}~\cite{FIRL}, \texttt{SVMIX}~\cite{SVMIX}, and \texttt{SAC}~\cite{SAC}. The experimental validation is conducted using the SMARTS~\cite{zhou2020smarts} simulation platform, specifically employing the challenging ``Unprotected Signal-Free Intersection'' scenario within a closed-loop ``Figure Eight" traffic configuration, which is a representative mixed-autonomy environment depicted in Fig. \ref{fig3}.

Our experimental configuration features a controlled cohort of $I=8$ CAVs governed by \texttt{MA-PMBRL}, coexisting with $I_\text{hv}$ HVs whose stochastic behavior is managed by the SMARTS environment's baseline intelligent controllers. All CAVs initiate navigation from a unidirectional approach lane, continuously traversing the intersection while optimizing for collision avoidance, congestion mitigation, and traffic flow efficiency. The vehicular interactions strictly adhere to the MDP formulation detailed in Section \ref{sec: preliminaries}, with specific state-action-reward mappings tailored for the ``Unprotected Intersection'' scenario as follows:

\begin{itemize}
\item \emph{Observable State Space}: Each CAV maintains a partial observation field limited to its kinematic status and proximate vehicular context. As described in Section \ref{sec:systemModel}, the perceptual state for agent $i$ at time $t$ comprises:
\begin{align*}
s_{t}^{(i)} = \Big( \underbrace{v_t^{(i)}, x_t^{(i)}, y_t^{(i)}}_{\substack{\text{ego}\\ \text{kinematics}}}, \underbrace{v_{t,a}^{(i)}, l_{t,a}^{(i)}}_{\substack{\text{leading}\\ \text{vehicle}}}, \underbrace{v_{t,e}^{(i)}, l_{t,e}^{(i)}}_{\substack{\text{following}\\ \text{vehicle}}}, b_t^{(i)} \Big) \\\in \mathbb{R}^8 \times \left\{0,1\right\}.
\end{align*}
Notably, $b_t^{(i)}$ functions as a collision risk flag triggered when any vehicle enters the emergency braking radius ($\left\|l_{t, a / e}\right\|<l_{\mathrm{safe}}$).
\item \emph{Control Actions}: The action space for each CAV integrates longitudinal control with lane change decisions:
\begin{align*}
    a_t^{(i)} = \Big( \underbrace{v_t^{(i)}}_{\substack{\text{target speed} \\ \text{(continuous)}}}, \underbrace{c_t^{(i)}}_{\substack{\text{lane change} \\ \text{(binary)}}}\Big) \in \mathbb{R} \times \{0,1\}
\end{align*}
The lane change dimension $c_t^{(i)}$ becomes inactive for single-lane configurations, reducing the action space to pure velocity regulation.

\item \emph{Cooperative Reward Design}: The reward function is designed to encourage collision-free velocity maximization while implicitly coordinating traffic flow through observations of neighboring vehicle velocities. Specifically, it is formulated as:
\begin{align}
\label{eq:rewardfunction}
    r_t^{(i)} = \underbrace{v_t^{(i)}}_{\substack{\text{ego velocity} \\ \text{optimization}}} + \lambda\big
    ( \underbrace{v_{t,a}^{(i)} + v_{t,e}^{(i)}}_{\substack{\text{neighbor vehicle} \\ \text{velocity coordination}}} \big) - \zeta \cdot \mathbb{I}(b_t^{(i)})
\end{align}
where $\lambda$ regulates the influence of neighboring vehicle velocities to facilitate coordinated traffic flow, and $\zeta$ penalizes emergency braking events to enhance both individual efficiency and emergent collective behavior. By incorporating observable neighbor velocities, the reward function fosters a balance between local optimization and global coordination.
\end{itemize}  

Given that directly summing rewards across multiple agents as defined in Eq. \eqref{eq:rewardfunction} may introduce computational redundancy and evaluation bias due to overlapping vehicle states, we propose a novel system-level utility metric grounded in traffic efficiency analysis. By incorporating the cumulative travel distance ${\Lambda}x_t^{(i)}$ of vehicle $i$ until the last time-step of episode $k$, the system utility metric is formulated as: 
\begin{align}
\label{eq:utilityMetric}
    {\rm Utility}\left(K\right)=\mathbb{E}_{\pi \sim \tilde{T}_{\phi_k}}\frac{1}{HI} \sum\nolimits_{i=1}^I \sum\nolimits_{t=0}^{H} {\Lambda}x_t^{(i)}
\end{align} 
where $\tilde{T}_{\phi_k}$ ($\forall k\in[K]$) denotes the updated dynamics model until the episode $k$ with $20$ total training episodes and $H=1500$ maximum collision-free time-steps per episode in single-lane scenarios. This metric mitigates state redundancy in naive reward aggregation while quantifying global traffic efficiency through normalized cumulative travel distances, ensuring robust multi-vehicle coordination evaluation. Finally, all experiments are conducted on the NVIDIA GeForce RTX 4090 with $5$ independent simulations, with full parameter specifications for the traffic environment, DNN architectures, and \texttt{MA-PMBRL} hyperparameters detailed in Table \ref{tb: setting}.  

\begin{table}[t!]
    \centering
    \caption{Parameter Configurations for the SMARTS Simulator and MA-PMBRL Framework.}
    \begin{center}
    \label{tb: setting}
    \begin{threeparttable} 
        \begin{tabular}{m{4.2cm} | m{3.3cm} } 
        \hline 
        \textbf{Parameter} & \textbf{Value} \tabularnewline
        \hline 
        \multicolumn{2}{c}{\textbf{SMARTS Simulator}} \tabularnewline
        \hline
        Discrete-time step  &  $\tau =0.1 \mathrm{~s}$ \tabularnewline
        No. of vehicles  & $I=8$ \tabularnewline
        No. of human-driven vehicles & $I_\text{hv} \in [4,12]$ \tabularnewline
        "Figure Eight" track length & $480\mathrm{~m}$ \tabularnewline
        Speed limits (min, max) & $0 \leq v \leq 13.89$ m/s \tabularnewline
        \hline 
        \multicolumn{2}{c}{\textbf{MA-PMBRL Framework}} \tabularnewline
        \hline
        Velocity observation & $v_t \in [0,13.89]$ m/s \tabularnewline
        Position coordinates & $z_t \in \mathbb{R}^2$ \tabularnewline
        Target velocity & $\bar{v}_{t} \in [0,13.89]$ m/s \tabularnewline
        Speed of nearby vehicles & $v_{t,a},v_{t,e} \in [0,13.89]$ m/s \tabularnewline
        Distance to nearby vehicles & $l_{t,a},l_{t,e}\in [0,75]$ m \tabularnewline
        Emergency braking indicator & $b^{(i)} \in \{0,1\}$ \tabularnewline
        Travel distance per vehicle & $\Lambda x_t^{(i)} \in \mathbb{R}$ \tabularnewline
        No. of episodes & $K=20$ \tabularnewline
        Max episode length & $H=1500$ \tabularnewline
        Communication range & $d \in [0,200]$ m \tabularnewline
        Model rollout length & $L_{\text{rollout}}=8$ \tabularnewline
        Reward weight & $\lambda=0.85$ \tabularnewline
        Safety penalty & $\zeta=7.5$ \tabularnewline
        \hline 
        \multicolumn{2}{c}{\textbf{Learning Rates}} \tabularnewline
        \hline
        PGD  & $\alpha_{\text{PGD}}=0.001$ \tabularnewline
        Actor network  & $\alpha_{\text{actor}}=0.0001$ \tabularnewline
        Critic network  & $\alpha_{\text{critic}}=0.001$ \tabularnewline
        Temperature parameter  & $\alpha_{\text{temp}}=0.001$  \tabularnewline
        \hline 
        \multicolumn{2}{c}{\textbf{Training Parameters}} \tabularnewline
        \hline
        Real-to-model data ratio & $0.7$ \tabularnewline
        Discount factor & $\gamma=0.98$ \tabularnewline
        Soft update parameter & $0.01$ \tabularnewline
        Random seed & $5$ \tabularnewline
        \hline 
        \multicolumn{2}{c}{\textbf{Probabilistic Deep Neural Network}} \tabularnewline
        \hline
        Replay buffer size & $20,000$ \tabularnewline
        Model buffer size  & $7,000$ \tabularnewline
        No. of hidden layers & $3$ \tabularnewline
        No. of hidden units & $256$ \tabularnewline
        Training batch size  & $1,000$ \tabularnewline
        Mini-batch size  & $512$ \tabularnewline
        Optimizer & Adam \tabularnewline
        \hline
        \end{tabular}
        \begin{tablenotes}
            \footnotesize
            \item[$\ast$] Parameters specific to \texttt{MA-PMBRL}.
            \item[$\dagger$] Temperature parameter ($\alpha_{\text{temp}}$) controls entropy regularization in maximum entropy RL.
        \end{tablenotes}
    \end{threeparttable}
    \end{center}
\end{table}

\subsection{Numerical Results}
\label{sec: numerical results}
\subsubsection{Validation of Performance Superiority}
\begin{figure}
	\centering
	\includegraphics[width=0.95\linewidth]{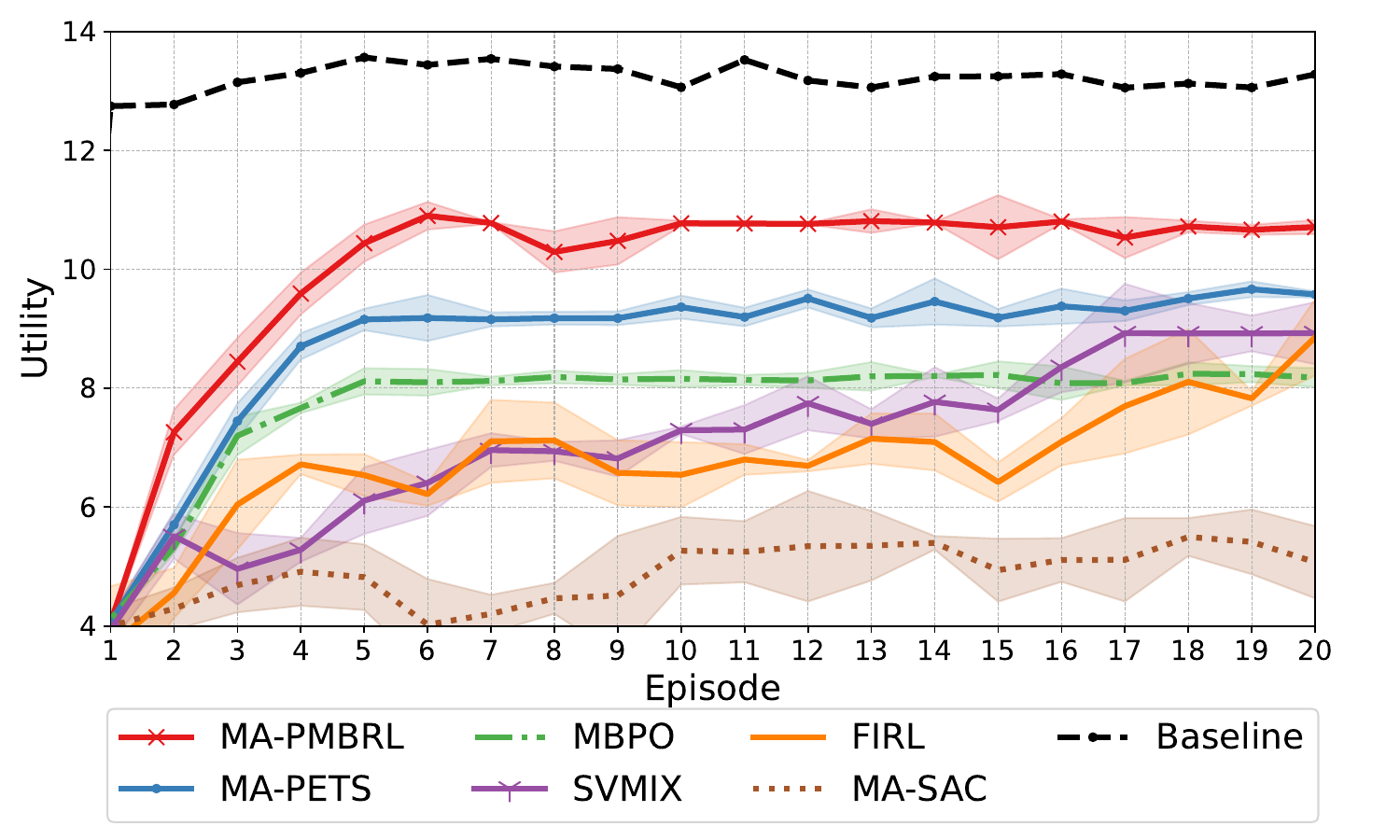}
	\caption{Comparison of utility in the single-lane ``Unprotected Intersection'' scenario.}
    \vspace{-5pt}
	\label{utility_single_lane}
\end{figure}

We evaluate the performance of \texttt{MA-PMBRL} for CAVs with $I =8$ and communication range $d=100$, comparing it with other state-of-art MBRL and MFRL algorithms such as \texttt{MA-PETS}, \texttt{MBPO}, \texttt{SVMIX}, \texttt{MA-SAC} and \texttt{FIRL} in Fig. \ref{utility_single_lane}. Furthermore, an optimal baseline is established using SMARTS, where all vehicles are controlled based on a typical car-following model that incorporates significant prior knowledge. It can be observed that \texttt{MA-PMBRL} achieves superior utility over the episodes, in terms of faster convergence and more consistent improvements throughout the training process. Specifically, the results demonstrate that \texttt{MA-PMBRL} outperforms most algorithms and maintains higher utility, leading to more efficient multi-agent coordination. This makes \texttt{MA-PMBRL} an effective choice for environments requiring high levels of coordination and performance stability.

\subsubsection{Impact of the Choice of $\phi$ in Eq. \eqref{eq: choose_phi}}
\begin{figure}
	\centering
	\includegraphics[width=0.95\linewidth]{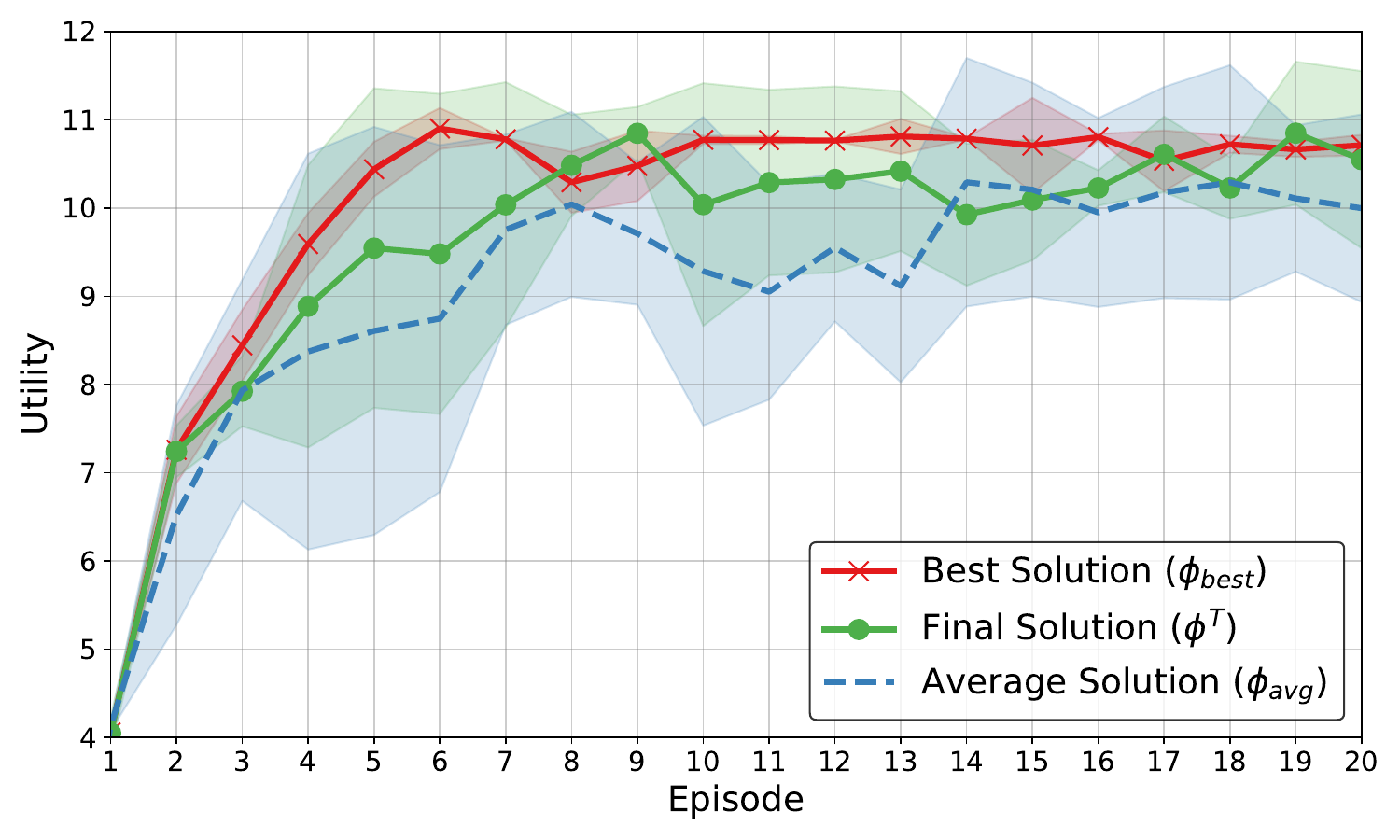}
	\caption{Comparison of utility of different choices of PGD in the single-lane ``Unprotected Intersection'' scenario.}
    \vspace{-5pt}
	\label{PGD_choice}
\end{figure}

We evaluate the effect and robustness of our choice of $\phi$ in PGD
, by comparing with the selection of 
the average solution $\phi_\mathrm{avg}$ 
and the final solution $\phi_X$. As shown in Fig. \ref{PGD_choice}, selecting the best solution ($\phi_\mathrm{best}$) consistently yields the highest utility across episodes and ensures the fastest convergence to optimal utility, making it the most effective strategy. In contrast, the final solution ($\phi_X$) exhibits a slower increase in utility than the best solution and eventually stabilizes at a slightly lower performance level. This could be due to its potential sensitivity to abrupt state changes or insufficient convergence, especially in dynamic or unstable network conditions. Therefore, it may fail to capture the optimal policy, leading to higher variance consistently. On the other hand, the average solution ($\phi_\mathrm{avg}$) may overlook some local optima, especially under favorable vehicle dynamics or network conditions, as these optimal solutions could be 'smoothed out' or diluted by the averaging process. As a result, while gradually improving, it consistently underperforms both the best and final solution strategies. These findings indicate that consistently selecting the best solution outperforms the other strategies, offering higher utility and faster convergence, making it the most effective choice for enhancing agent performance in the ``Unprotected Intersection'' scenario.
  
\subsubsection{Impact of Different Communication Ranges $d$}
\begin{figure}
	\centering
	\includegraphics[width=0.95\linewidth]{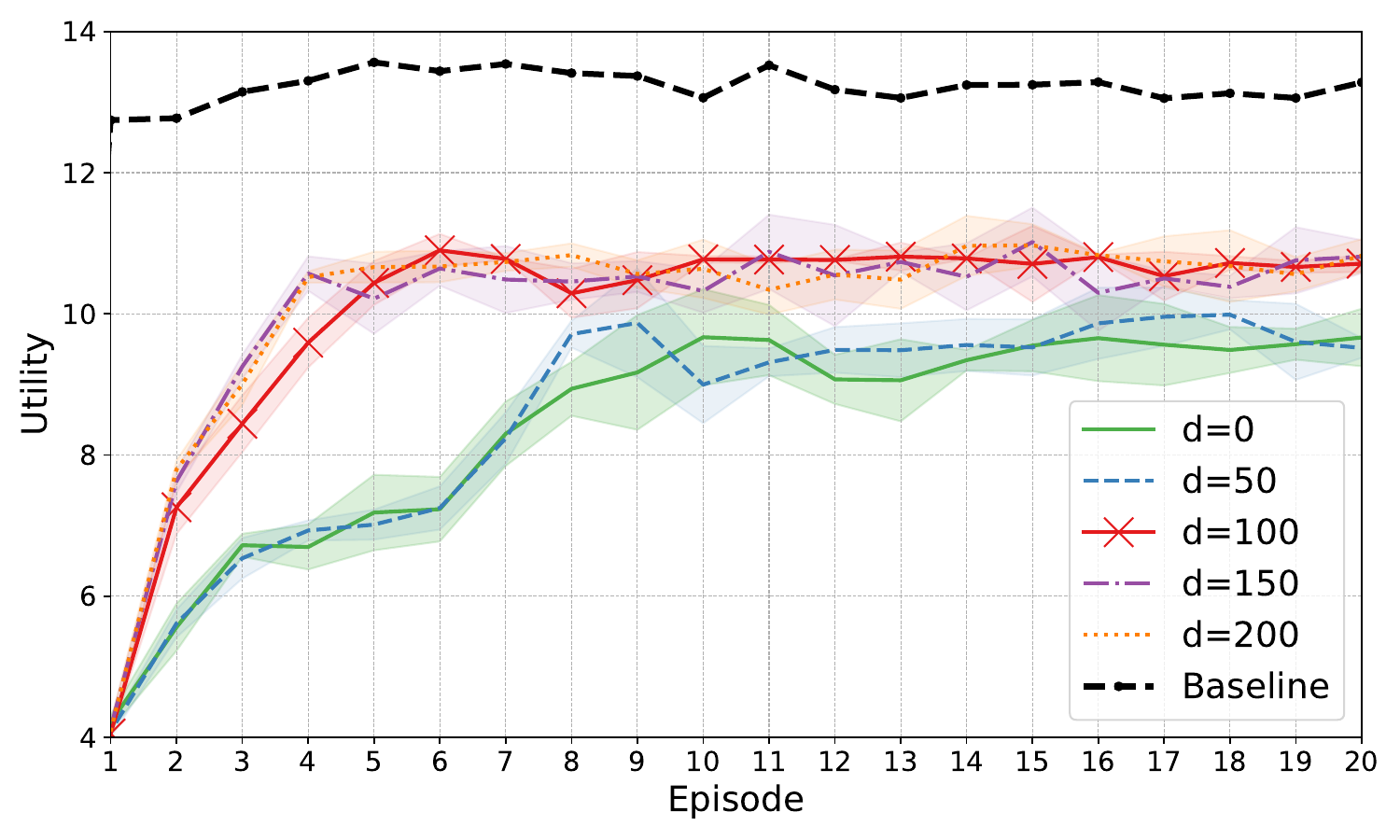}
	\caption{Performance comparison under different communication range $d$.}
    \vspace{-5pt}
	\label{d_utility_single_lane}
\end{figure}

\begin{figure}
	\centering
	\includegraphics[width=0.95\linewidth]{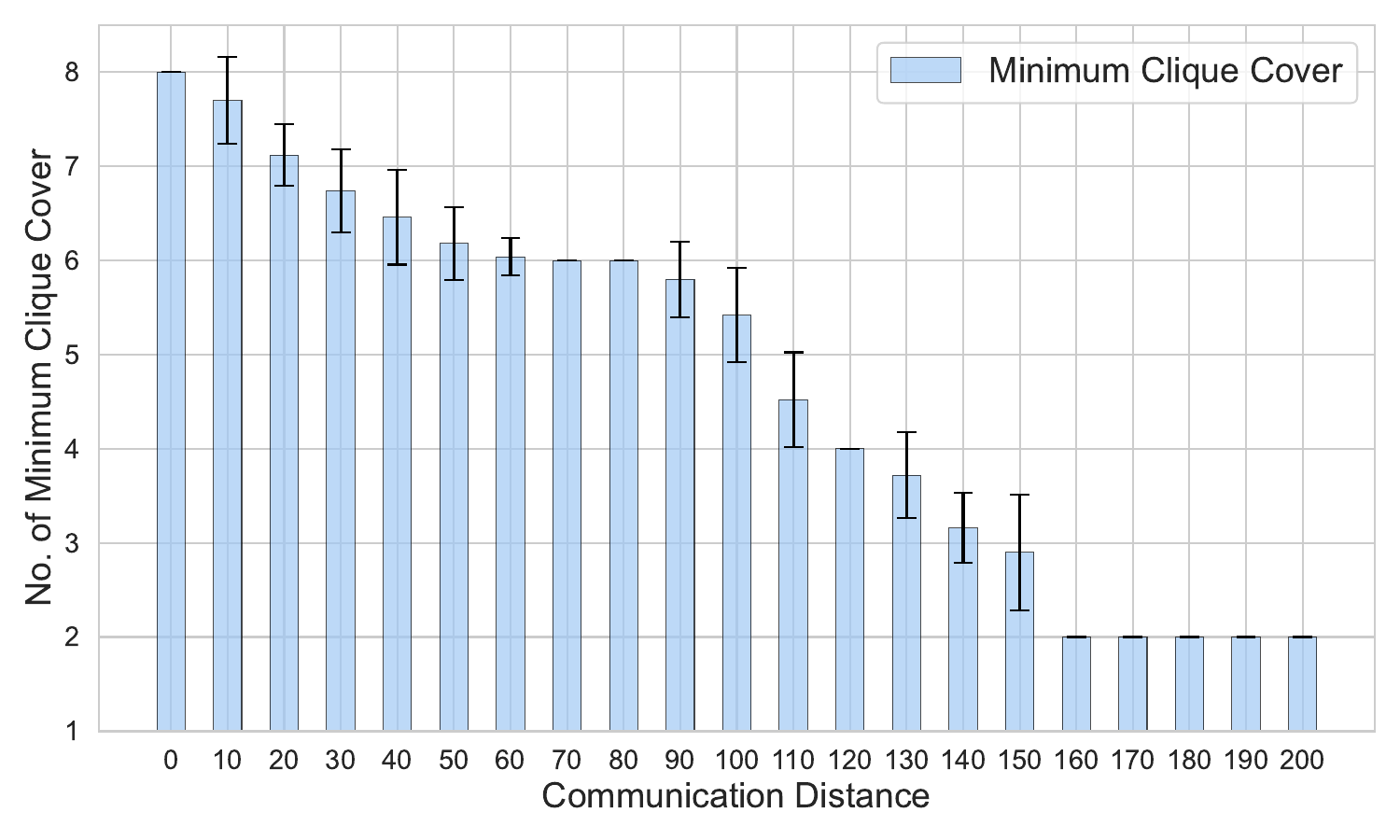}
	\caption{Correlation between number of minimum clique cover $\bar{\chi}\left(\mathcal{G}_{d,k}\right)$ and communication distance $d$.}
    \vspace{-5pt}
	\label{clique_cover}
\end{figure}

We investigate 
the performance of \texttt{MA-PMBRL} by varying $d$ from $0$ to $200$, as presented in Fig. \ref{d_utility_single_lane} and Table \ref{tb: communications overhead}. As expected, Table \ref{tb: communications overhead} confirms that increasing the communication range significantly raises the communication overhead. Meanwhile, as illustrated in Fig. \ref{d_utility_single_lane}, an extended communication range enhances learning efficiency, leading to notable improvements in agility and safety. However, Fig. \ref{d_utility_single_lane} also reveals that once the communication range surpasses a certain threshold, further expansion provides minimal additional benefits to policy performance, whereas Table \ref{tb: communications overhead} demonstrates an exponential growth in communication overhead. This observation suggests a trade-off between learning performance and communication efficiency. To further explore this interplay, we examine the relationship between the minimum number of clique covers $\bar{\chi}\left(\mathcal{G}_{d,k}\right)$ and the communication distance $d$ across $20$ independent trials. As depicted in Fig. \ref{clique_cover}, while the minimum number of clique covers stabilizes beyond a specific communication range threshold, a general inverse correlation exists between the number of clique covers and the communication range. This finding aligns with our theoretical results and the experimental insights obtained from Fig. \ref{d_utility_single_lane}, reinforcing the necessity of balancing communication efficiency and learning effectiveness in multi-agent coordination.

\begin{table}[t]
    	\centering
            \caption{\centering{The communication overheads of different communication range $d$.}}
        \label{tb: communications overhead}
	    \begin{tabular}{m{2cm}|ccccc}
            \toprule
            \textbf{Name} & \multicolumn{5}{c}{\textbf{Value}}\\
	    	\midrule
	    	Communication Distance $ d $ &$0$& $50$& $100$& $150$& $200$\\
	    	\hline  
	    	Communication Overheads &$0$& $241.9$& $1,443.7$& $5,635.3$& $7,192.5$\\
	    	\bottomrule
	    \end{tabular}
    \end{table}	

\subsubsection{Impact of Rollout Length $L_{\text{rollout}}$}
\begin{figure}
	\centering
	\includegraphics[width=0.95\linewidth]{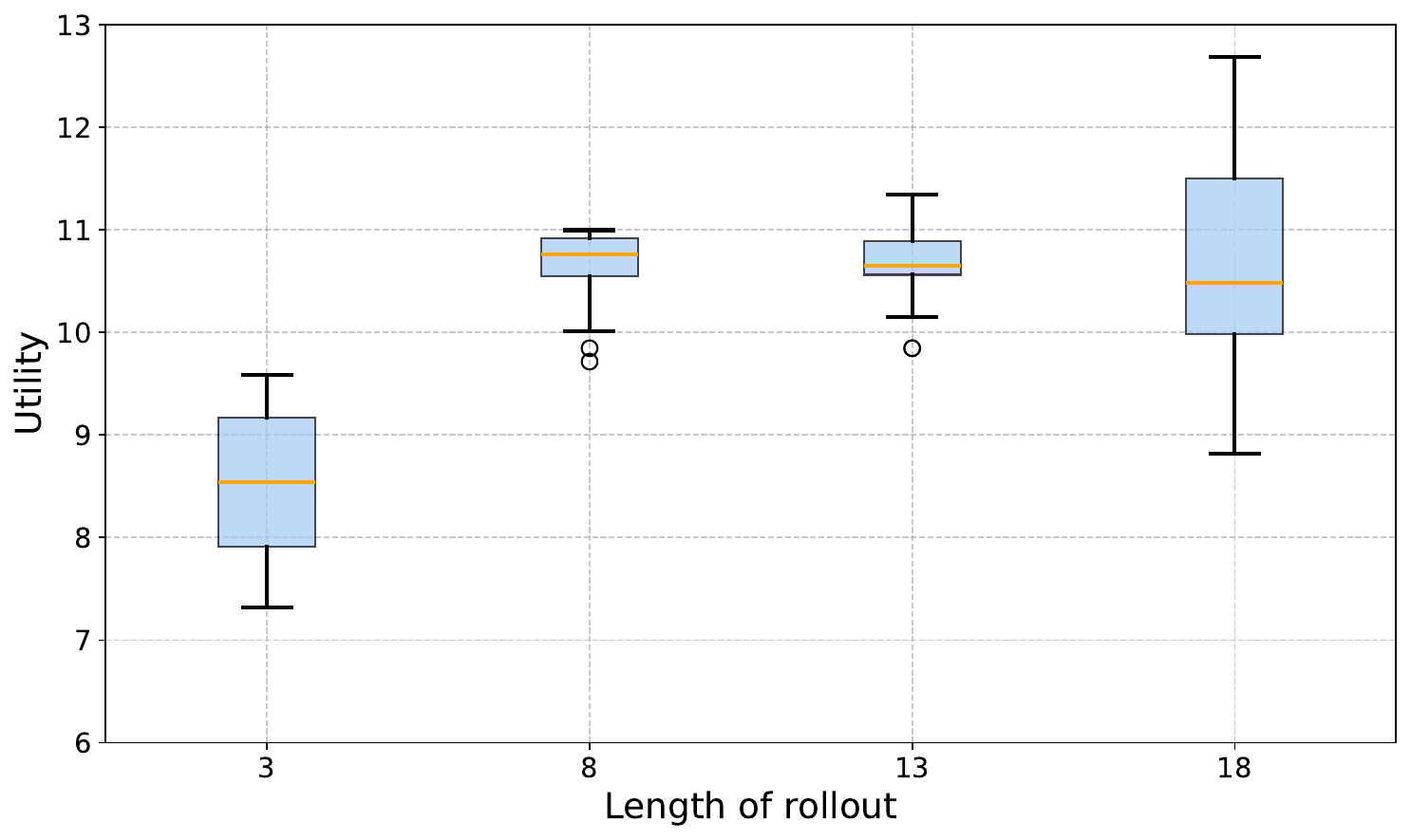}
	\caption{Comparison of utility across different rollout lengths.}
    \vspace{-5pt}
	\label{rollout_box}
\end{figure}

Given the significance of rollout length to decision-making quality and overall utility performance, we experiment with different optimal rollout lengths, while presenting utility across the last five episodes after convergence. As shown in Fig. \ref{rollout_box}, the utility varies dramatically along with the variations in rollout length. Specifically, the shortest rollout length corresponds to lower utility and higher variability, suggesting that shorter rollouts lack sufficient information for accurate predictions. As the rollout length increases to $8$ or $13$, the utility improves with decreased variability, indicating that a moderate length allows for more informed, stable decisions. 
However, at a length of $18$, the utility peaks but exhibits the greatest variability, indicating that longer planning horizons may improve optimization at the cost of stability. In conclusion, longer rollout lengths generally improve utility by accounting for more future states, but beyond a certain point, additional length offers diminishing returns. Considering that longer rollouts incur higher computational and storage costs, a length of $8$ is defaultly chosen for the ``Unprotected Intersection'' scenario.

\section{Conclusion and Discussion}
\label{sec: conclusion}
In this paper, we have studied a fully decentralized MBRL-based self-decision control for CAVs with communication constraints and limited datasets with partial coverage. In particular, we have proposed the \texttt{MA-PMBRL} algorithm with a significant performance improvement in terms of sample efficiency. Specifically, \texttt{MA-PMBRL} facilitates the efficient exchange of collected samples among individual agents and their neighboring vehicles within a specified communication range. In addition, it incorporates a pessimistic model-based min-max framework, which ensures vehicle safety by mitigating the reliance on overly subjective uncertainty estimations. Furthermore, \texttt{MA-PMBRL} leverages PGD within the critic network, which greatly simplifies the computational process of deriving effective control policies from pessimistic estimations. Afterward, we have derived a theoretical analysis of \texttt{MA-PMBRL} under communication constraints and partial coverage, building on the theoretical foundation of graph theory and concentration inequalities. We have validated the superiority of \texttt{MA-PMBRL} over classical MFRL and MBRL algorithms and demonstrated the contribution of communication to multi-agent MBRL.

There are many interesting directions to be explored in the future. For example, although \texttt{MA-PMBRL} achieves rapid convergence and demonstrates superior performance compared to many existing methods, a noticeable performance gap remains relative to the optimal baseline within certain scenarios. This discrepancy primarily stems from the inherent sensitivity of model-based approaches to dataset quality. Nevertheless, the exceptional sample efficiency and rapid convergence properties of MBRL make it particularly suitable as a pre-training method, offering valuable initialization guidance to MFRL algorithms during the early stages of training when data availability is severely limited. Furthermore, the current study does not explicitly address the prevalent issue of OOD data encountered in MBRL frameworks, which may constitute another factor that limits the performance. Finally, the theoretical guarantees established in this paper are derived explicitly within the framework of the employed SAC algorithm; future research may explore extending the PAC guarantee to more general RL settings.


\newpage
\begin{appendices}
	\section{Useful Lemmas}

Lemma \ref{lem: Policy_improvement} below shows that the difference in values between two policies at a given state $s$ can be interpreted as the expected sum of discounted advantages of one policy over the other along the trajectory generated by $\pi$.
\begin{lemma}[Policy improvement of Lemma 6.1 of \cite{Kakade2002ApproximatelyOA}]
\label{lem: Policy_improvement}
For two policies $\pi$ and $\pi_t$, with initial state $s$ and trajectories $\tau \sim T_{\pi}(\tau | s_0 = s)$ under policy $\pi$, the difference in value functions $V_{\pi}(s) - V_{\pi_t}(s)$ can be expressed as the expected cumulative advantage of $\pi$ over $\pi_t$ along the trajectory:
\begin{align}
    V_{\pi}(s) - V_{\pi_t}(s) &= \mathbb{E}_{\tau \sim T_{\pi}(\tau | s_0 = s)}\left[\sum_{t=0}^{\infty} \gamma^t A_{\pi_t}(s_t, a_t)\right]\nonumber\\
    &= \frac{1}{1 - \gamma} \mathbb{E}_{\tau \sim T_{\pi}(\tau | s_0 = s)}\left[Q_{\pi_t}(s, a) - V_{\pi_t}(s)\right], \nonumber
\end{align}
where $Q_{\pi_t}$ is the state-action value function, and $V_{\pi_t}$ is the state value function. The advantage function $A_{\pi_t}(s_t, a_t)$ quantifies the difference between $Q_{\pi_t}$ and $V_{\pi_t}$, highlighting the relative benefit of taking action $a_t$ at state $s_t$ under policy $\pi_t$ .
\end{lemma}

To establish a theoretical foundation for estimating model accuracy, the following lemma provides an MLE guarantee that bounds the discrepancy between the learned and true environment models, forming the basis for evaluating distributional accuracy.
\begin{lemma}[MLE guarantee from Section E of \cite{agarwal2020optimality}] 
\label{lem: MLE_guarantee}
Given a set of models $\mathcal{M}_{\mathcal{D}}=\{T_\phi: \mathcal{S} \times \mathcal{A} \rightarrow \Delta(\mathcal{S}\times \mathcal{R})\}$ with $T \in \mathcal{M}_{\mathcal{D}}$, and a dataset $\mathcal{D}=\left\{s_t, a_t, s_{t+1}, r_t\right\}_{t=1}^n$ with $s_t, a_t \sim \rho$, and $s_{t+1},  r_t \sim T\left(\cdot |s_t, a_t\right)$, let $\tilde{T}_{\mathrm{MLE}}$ be:
\begin{align}
    \tilde{T}_{\mathrm{MLE}}=\underset{T_\phi \in \mathcal{M}_{\mathcal{D}}}{\arg \min } \sum_{t=1}^n-\ln T_\phi\left(s_{t+1}, r_t \mid s_t, a_t\right).\nonumber
\end{align}
With probability at least $1-\delta$, we have:
\begin{align}
    \mathbb{E}_{s, a \sim \rho} \operatorname{TV}\left(\tilde{T}_{\mathrm{MLE}}(\cdot \mid s, a), T(\cdot \mid s, a)\right)^2 \lesssim \frac{\ln \left(\mid \mathcal{M}_{\mathcal{D}} / \delta \mid\right)}{n}.\nonumber
\end{align}
\end{lemma}
\begin{lemma}[Lemma 7 of \cite{hongpessimistic}]
\label{lem: generalized_simulation}
Suppose $\mathcal{S}$, $\mathcal{A}$, $r$, $\gamma$, and $\mu_0$ are fixed, where the state space $\mathcal{S}$ and action space $\mathcal{A}$ can be infinite sets and $r:\mathcal{S}\rightarrow \mathbb{R}$ is any real-valued reward function. Given two arbitrary models $T$ and $\tilde{T}$, and any policy $\pi: \mathcal{S}\rightarrow \Delta(\mathcal{A})$, If the value function $V_{\tilde{T}}^\pi(s)$ is bounded, i.e., $-C\leq V_{\tilde{T}}^\pi(s)\leq C$ for all $s\in\mathcal{S}$, then we have:
\begin{align}
    \left|V_T^\pi - V_{\tilde{T}}^\pi\right| \leq \frac{2C\gamma}{1-\gamma}\mathbb{E}_{(s,a)\sim d_T^\pi}\left[\mathrm{TV}\left(T(\cdot|s,a),\tilde{T}(\cdot|s,a)\right)\right].\nonumber
\end{align}
\end{lemma}

\begin{lemma}[Boundedness of Discounted Return]
\label{lem: Boundedness_Q}
Suppose that the immediate reward function of the environment $\mathcal{R}(s, a)$ satisfies $|\mathcal{R}(s, a)| \leq R_{\max}$ for all pairs of state action $(s, a)$, and let the discount factor $\gamma \in[0,1$). Then, the cumulative discounted return (Q-value) is bounded as follows:
\begin{align}
    |Q(s, a)| \leq \frac{R_{\max}}{1-\gamma}.\nonumber
\end{align}

\begin{lemma}[Distribution Conversion Lemma of \cite{JMLR:v22:19-736}]
\label{lem: Distribution_Conversion_Lemma}
    Let $d^{\pi_1}$ and $d^{\pi_2}$ be the state distributions induced by two policies $\pi_1$ and $\pi_2$, respectively, in a Markov decision process. Assume $d^{\pi_1} \ll d^{\pi_2}$ (i.e., $d^{\pi_1}$ is absolutely continuous with respect to $d^{\pi_2}$ ). For any measurable function $f(s)$, the following holds:
    \begin{align}
        \mathbb{E}_{s \sim d^{\pi_1}}[f(s)]=\mathbb{E}_{s \sim d^{\pi_2}}\left[f(s) \frac{d^{\pi_1}(s)}{d^{\pi_2}(s)}\right]. \nonumber
    \end{align}
\end{lemma}

\end{lemma}
The following two lemmas are useful properties for the log-linear parametric class (defined in Assumption \ref{as:log_linear_policy}).
\begin{lemma}[Lemma 8 of \cite{hongpessimistic}]
\label{lem: log_linear_1}
For any policy $\pi_\psi$ in the log-linear parametric class, the following holds for any $(s, a)$:
\begin{align}
    \nabla_\psi \log \pi_\psi(a \mid s)=\varphi_{s, a}-\mathbb{E}_{a^{\prime} \sim \pi_\psi(\cdot \mid s)} \varphi_{s, a^{\prime}}.\nonumber
\end{align}
\end{lemma}

\begin{lemma}[Lemma 9 of \cite{hongpessimistic}]
\label{lem: log_linear_2}
For any policy $\pi_\psi$ within the log-linear parametric class, if the feature vector satisfies $\left\|\varphi_{s, a}\right\|_2 \leq \varphi_{\max}$, then $\log \pi_\psi(a \mid s)$ is a $\varphi_{\max}^2$-smooth function of the parameter$\psi$ for all $(s, a)$. Formally, we have:
\begin{align}
    &\left\|\nabla_\psi \log \pi_{\psi_1}(a \mid s)-\nabla_\psi \log \pi_{\psi_2}(a \mid s)\right\|_2 \nonumber\\
    &\qquad\leq \varphi_{\max}^2\left\|\psi_1-\psi_2\right\|_2\nonumber
\end{align}
for any s, a, $\psi_1, \psi_2$.
\end{lemma}

\begin{lemma}[Growth Inequality for $\beta$-Smoothness (Section 2.1 of \cite{nesterov2013introductory}) and Hessian Boundedness (Theorem 5.12 of \cite{Smoothness})]
\label{lem: smoothness_property}
Let $f: \mathbb{R}^n \rightarrow \mathbb{R}$ be a differentiable function. If $f$ is $\beta$-smooth, then its Hessian matrix (i.e., the second-order derivative) at any point $x$ is bounded by $\beta I$, meaning that $-\beta I \preceq \nabla^2 f(x) \preceq \beta I$. Here, $I$ denotes the identity matrix, and $\beta$ is a non-negative constant. 
The deviation from the first-order approximation is bounded by:
\begin{align}
    |f(y) -f(x)-\nabla f(x)^T(y-x)|\leq \frac{\beta}{2}\|y-x\|^2.\nonumber
\end{align}

\end{lemma}

\begin{lemma}[Eq. (6) of \cite{9867146}]
\label{lemma:clique_number}
Let $\mathcal{C}_{d,k}$ represent the clique covering of the graph $\mathcal{G}_{d,k}$, where the graph $\mathcal{G}_{d,k}$ consists of $I$ nodes, and the cliques within $\mathcal{C}_{d,k}$ have node counts $|C_{C \in \mathcal{C}_{d,k}}|$. The minimum clique cover $\bar{\chi}\left(\mathcal{G}_{d,k}\right)$ maintains a consistent relationship  
 $\sum_{C \in \mathcal{C}_{d,k}}\sqrt{|C|} =\sqrt{\bar{\chi}\left(\mathcal{G}_d\right)I}$.
\end{lemma}

\section{Bound of $\Delta_a^{(i)}$ and $\Delta_c^{(i)}$}
\label{sec:bound_delta_a_c}
To rigorously establish a bound on $\Delta_a^{(i)}$ for each vehicle $i$, we proceed by deriving a uniform upper bound on the quantity $ V_{T,\pi^{\ast}}^{(i)} - \min_{T_\phi \in \mathcal{M}_{\mathcal{D}}} V_{T_\phi,\pi^{\ast}}^{(i)}$, which encapsulates the discrepancy between the value function under the optimal model $T$ and its minimum achievable counterpart across the model set $\mathcal{M}_{\mathcal{D}}$. This upper bound is formally provided in Lemma \ref{th: pac_cppo} and is supported by applying Lemma \ref{lem: MLE_guarantee}, ensuring a consistent measure of deviation applicable across all agents \cite{uehara2022pessimistic}. Consequently, we can directly infer that
\begin{align}
    \Delta_a^{(i)} &= \frac{1}{H+1} \sum_{t=0}^H \left( V_{T,\pi^{\ast}}^{(i)} - \min_{T_\phi \in \mathcal{M}_{\mathcal{D}}} V_{T_\phi,\pi^{\ast}}^{(i)} \right) \nonumber \\
    &\leq c_1(1-\gamma)^{-2} \sqrt{C_{\pi^{\ast}}\, \Lambda_{(s, a) \sim \rho,\max}} \nonumber \\
    &\leq c_1 (1-\gamma)^{-2} \sqrt{\frac{C_{\pi^{\ast}} \ln \left( c_2 \left|\mathcal{M}_{\mathcal{D}}\right| / \delta \right)}{|\mathcal{D}_E^{(i)}|}}.\nonumber
\end{align}
where this bound incorporates both the discount factor $\gamma$ and the maximum TV distance $\Lambda_{(s, a) \sim \rho,\max}$ between the true MDP transition function and the estimated MDP transition function under $(s, a) \sim \rho$, thus providing a robust upper limit for $\Delta_a^{(i)}$ across all horizon steps.

To bound $\Delta_c^{(i)}$, we assume that $T \in \mathcal{M}_{\mathcal{D}}$ holds with high probability, as demonstrated in Appendix E of \cite{uehara2022pessimistic}. Under this assumption, the critic step in Algorithm \ref{al: MA-PMBRL} ensures that $ \min_{T_\phi \in \mathcal{M}_{\mathcal{D}}} V_{T_\phi,\pi_t}^{(i)}$ achieves the minimum value among all models in $\mathcal{M}_{\mathcal{D}}$, leading to the inequality $\min_{T_\phi \in \mathcal{M}_{\mathcal{D}}} V_{T_\phi,\pi_t}^{(i)} \leq V_{T,\pi_t}^{(i)}, \forall t \in [H],$ which directly implies
\begin{align}
    \Delta_c^{(i)}=\frac{1}{H+1} \sum_{t=0}^H \left( \min_{T_\phi \in \mathcal{M}_{\mathcal{D}}} V_{T_\phi,\pi_t}^{(i)} - V_{T,\pi_t}^{(i)} \right) \leq 0.\nonumber
\end{align}
Thus, $\Delta_c^{(i)}$ contributes no positive deviation, as it is non-positive by construction, serving as a stabilizing term in the overall bound.


\end{appendices}
\bibliographystyle{IEEEtran}
\bibliography{bib1}
\end{document}